\documentstyle[sprocl]{article}
\def\rsim{\lower0.6ex\vbox{\hbox{$ \buildrel{\sim}\over{ >}\ $}}}

\def\be{\begin{equation}}
\def\ee{\end{equation}}
\def\bea{\begin{eqnarray}}
\def\eea{\end{eqnarray}}

\def\tensorD{\widetilde D}
\def\boldtensorD{\widetilde {\bf D}}
\begin{document}

\title{INTRODUCTION TO THE NRQCD FACTORIZATION APPROACH TO HEAVY 
	QUARKONIUM\footnote{based on a series of lectures presented at the
	Third International Workshop on Particle Physics Phenomenology in 
	Taipei, November 1996.}}

\author{ E. Braaten }
\address{Department of Physics, Ohio State University,
        Columbus OH 43210 USA}

\maketitle
\abstracts{ I present an introduction to the NRQCD factorization method
for calculating annihilation rates and inclusive production
rates for heavy quarkonium.   Using this
method, annihilation decay rates and sufficiently inclusive cross sections
are factored into long-distance NRQCD matrix elements
and perturbative short-distance coefficients.
I derive the velocity-scaling rules that are used to estimate the
magnitudes of the nonperturbative NRQCD matrix elements and
I describe perturbative matching methods for
calculating their short-distance coefficients.
Some simple applications for which NRQCD factorization
methods have dramatic implications are discussed.
}

\section{Introduction}
\label{sec:Int}

The NRQCD factorization approach is a systematic framework
for analyzing annihilation decay rates and sufficiently inclusive production
cross sections for heavy quarkonium.\cite{B-B-L}  This method separates 
the effects of short distances that are comparable to or smaller than 
the inverse of the heavy quark mass from the effects of 
longer distance scales.  Short-distance effects are calculated
using perturbative QCD and
long-distance effects are described by matrix elements
in an effective field theory called {\it nonrelativistic QCD} (NRQCD).
This paper is an introduction to the NRQCD factorization approach.
In Section~\ref{sec:NRQCD}, the effective lagrangian
for NRQCD is constructed, the velocity-scaling rules that
are used to estimate the magnitude of NRQCD matrix elements are derived,
and perturbative matching methods for calculating
short-distance coefficients are described.
The application of NRQCD factorization methods
to annihilation rates and to inclusive production rates
are discussed in Sections~\ref{sec:AnnDec} and \ref{sec:IncProd}.

\section{Heavy Quarkonium and NRQCD}
\label{sec:NRQCD}

\subsection{Scales in Quarkonium Physics}
\label{sec:scales}

Heavy quarkonium is a meson containing a heavy quark and its antiquark.
There are a number of different energy scales that play an important role in
the quarkonium physics.
While the many scales make the physics
complex, they also make it interesting.
The scales of quarkonium include the mass $M$ of the heavy
quark, its typical momentum $M v$, and its typical kinetic energy $M v^2$.
The quark mass $M$ sets the total energy scale for annihilation decays
and the scale of the kinematic threshold for onium production.
The inverse of the typical momentum $Mv$ is the length scale
for the size of the onium state.  The typical kinetic energy $Mv^2$ is the
scale of the splittings between radial excitations and between
orbital-angular-momentum excitations in the onium spectrum.  For both
charmonium and bottomonium, the splittings between the two lowest $^3S_1$
states ($J/\psi$ and $\psi'$ for charmonium, $\Upsilon$ and $\Upsilon'$
for bottomonium) are approximately 600 MeV.  The splittings between the lowest
$^3S_1$ and $^3P_J$ states ($J/\psi$ and $\chi_{cJ}$ for charmonium,
$\Upsilon$ and $\chi_{bJ}$ for bottomonium) are approximately 400 MeV.  We
take 500 MeV as an estimate for the scale $Mv^2$.  In Table~\ref{tab:scales},
we give our estimates for the scales
$M$, $Mv$, and $Mv^2$ for charmonium and bottomonium.
The estimate for $M$ is half the mass of the lowest energy level of quarkonium.
The estimate for the scale $Mv$ is the geometric mean of the 
estimates for $M$ and $Mv^2$.  From 
Table~\ref{tab:scales}, we see that $v^2$ is approximately 1/3 for
charmonium and 1/9 for bottomonium.  These values are small enough to justify
a theoretical approach based on an expansion in powers of $v$.

\begin{table}[t]
\caption{Quarkonium energy scales \label{tab:scales}}
\vspace{0.2cm}
\begin{center}
\begin{tabular}{|l|ccr|}
\hline
&&&\\
& $c \bar c$ & $b \bar b$ & $t \bar t \qquad$ \\
\hline
$M$    & 1.5 GeV & 4.7 GeV & 180 GeV \\
$Mv$   & 0.9 GeV & 1.5 GeV &  16 GeV \\
$Mv^2$ & 0.5 GeV & 0.5 GeV & 1.5 GeV \\
\hline
\end{tabular}
\end{center} \end{table}

Another important energy scale in quarkonium physics is $\Lambda_{QCD}$, the
scale of nonperturbative effects involving gluons and light quarks.
The potential energy $V(R)$ between a quark and antiquark separated
by a distance $R$
varies from Coulombic at sufficiently short distances,
\begin{equation}
V(R) \;\approx\;  - {4 \over 3} {\alpha_s(1/R) \over R}
	\quad {\rm as} \; R \to 0,
\label{V-smallR}
\end{equation}
to linear at sufficiently
long distances:
\begin{equation}
V(R) \;\approx\;  \kappa^2 R \quad {\rm as} \; R \to \infty,
\label{V-largeR}
\end{equation}
where empirically $\kappa \simeq$ 450 MeV.  Since $\kappa$ is independent
of $M$,
it must be proportional to $\Lambda_{QCD}$.

To clarify the relation between the scale $\Lambda_{QCD}$ and the other scales
in quarkonium physics, we consider two limiting cases.
The first case is a heavy quark whose mass $M$ is
large enough that the onium wavefunction is dominated
by the Coulombic term in the potential.
The size of a bound state is
determined by a balance between the kinetic energy and the potential energy,
so we must have
\begin{equation}
Mv^2 \;\sim\; {4 \over 3} {\alpha_s(1/R) \over R}.
\label{balance:largeM}
\end{equation}
The size of the bound state is comparable to the inverse of the typical
momentum $Mv$ of the heavy quark in the bound state.
Setting $R \sim 1/(Mv)$ in (\ref{balance:largeM}), we obtain
\begin{equation}
v \;\sim\; \alpha_s (Mv) .
\label{v-estimate}
\end{equation}
This equation can be solved self-consistently for $v$ as a function of $M$.
If $M$ is sufficiently large, the resulting value of $Mv^2$
is much greater than $\Lambda_{QCD}$.
For the top quark with mass $M = 180$ GeV, we obtain $v \simeq 0.18$.
An alternative estimate for $v$, which is closer in spirit to the estimates 
used for charmonium and bottomonium in Table~\ref{tab:scales},
is obtained by taking $Mv^2$ to be the splitting between the ground state 
and the first excited state in the potential (\ref{V-smallR}).
This splitting is approximately 1.5 GeV, which corresponds to 
$v \simeq 0.09$.
This value for $Mv^2$ and the corresponding estimate for the scale $Mv$  
are included in Table~\ref{tab:scales}.
These would be the appropriate scales for toponium
if the top quark were stable
enough to form bound states.

The second limiting case is a heavy quark whose mass $M$ is
small enough that the wavefunction is dominated by the
linear term (\ref{V-largeR}) in the potential, but
still large enough that the onium is nonrelativistic.  The balance between the
kinetic and potential energies requires
\begin{equation}
M v^2 \;\sim\; \kappa^2 R.
\label{balance:smallM}
\end{equation}
Setting $R \sim 1/(Mv)$, we find that $\kappa \sim M v^{3/2}$.
Identifying $\kappa$ with the
scale $\Lambda_{QCD}$, we find that the typical velocity of the
heavy quark is such that
this scale is intermediate between the scales $Mv$ and $Mv^2$.
If the Coulombic and linear regions of the potential were both 
equally important, then the relations (\ref{balance:largeM}) 
and (\ref{balance:smallM}) would be satisfied simultaneously.  
The fact that these relations are compatible suggests that
the scaling relation (\ref{v-estimate}) might be applicable 
even for rather low values of $M$.

It is a remarkable coincidence of quarkonium physics
that the scales $Mv^2$ are almost identical for charmonium and bottomonium.
The fact that this scale is so insensitive to the value of $M$ suggests that 
$\Lambda_{QCD}$ should be identified with the scale $M v^2$ for 
quark masses in this range.  
This identification is supported by the numerical value of the scale $Mv^2$ 
given in Table~\ref{tab:scales}.

If any of the scales $M$,
$M v$, and $M v^2$ is large enough compared to $\Lambda_{QCD}$, then the
effects of that scale can be calculated using QCD perturbation theory.
The values of the running coupling constant of QCD at the scales
$M$, $M v$, and $M v^2$ are given in Table~\ref{tab:alphas}
for charmonium, bottomonium, and toponium.
All three scales are perturbative in the case of toponium.
For charmonium and bottomonium, $\alpha_s(M)$ is small enough to justify
perturbation theory at the scale $M$. The coupling constant $\alpha_s(Mv)$
is also small enough that perturbation theory seems reasonable. 
However, potential model calculations imply that the wavefunctions of 
charmonium and bottomonium have significant support in  the 
linear region of the potential, and this suggests that nonperturbative 
effects may be significant at the scale $Mv$.  As for the scale $M v^2$,
it is hopelessly nonperturbative for charmonium and bottomonium.

\begin{table}[t]
\caption{Value of the QCD coupling constant at the characteristic momentum
scales for heavy quarkonium
\label{tab:alphas}}
\vspace{0.2cm}
\begin{center}
\begin{tabular}{|l|ccr|}
\hline
&&&\\
& $c \bar c$ & $b \bar b$ & $t \bar t \;$ \\
\hline
$\alpha_s(M)$     & 0.35    & 0.22    & 0.11  \\
$\alpha_s(M v)$   & 0.52    & 0.35    & 0.16  \\
$\alpha_s(M v^2)$ & $\gg$ 1 & $\gg$ 1 & 0.35  \\
\hline
\end{tabular}
\end{center} \end{table}

In addition to the scales $M$, $M v$, $M v^2$, and $\Lambda_{QCD}$, there
are also kinematic energy scales that can play an important role in quarkonium
physics.  For example, in the production of quarkonium, the total
center-of-mass energy $\sqrt s$ and the transverse momentum $p_T$ of the onium
can be important.  Unravelling the effects of the various energy scales
is essential in order to understand quarkonium physics.
This is particularly important for charmonium and bottomonium,
because the coupling constant
$\alpha_s$ runs rather dramatically with the momentum at scales of order $M$
and smaller.

The NRQCD factorization approach is based on separating
short-distance effects involving momenta of order $M$ or larger from 
those effects that involve the smaller momentum scales $Mv$, $Mv^2$,
and $\Lambda_{QCD}$.
The scale $M$ is assumed to be perturbative, so that short-distance
effects can be calculated using perturbation expansions in $\alpha_s(M)$.
No assumption is made about the validity of perturbation theory
at the scale $Mv$.  Instead, we exploit the fact that in a
nonrelativistic bound state, the typical velocity $v$ provides a small
expansion parameter.

\subsection{Integrating out Relativistic Effects}
\label{sec:IntRel}

One way to separate the effects of the momentum scale $M$ from the lower
momentum scales in a field theory is to integrate out all modes with
momenta greater than some cutoff $\Lambda$ that is much less than $M$.  
The result of this
renormalization group transformation is a Wilsonian effective field theory
that describes the modes with momenta smaller than $\Lambda$.  
All effects of the
scale $M$ are encoded in the parameters of the effective field theory.
In our case, the original field theory is QCD with a heavy quark.
It is described by the lagrangian
\begin{equation}
{\cal L}_{\rm QCD} \;=\;
{\cal L}_{\rm light} + \overline \Psi (i \gamma^\mu D_\mu - m_Q) \Psi,
\label{L-QCD}
\end{equation}
where ${\cal L}_{\rm light}$ is the lagrangian that describe
gluons and light quarks.
The mass parameter $m_Q$ of the heavy quark can be identified with $M$.
It is implicit in the Lorentz-invariant lagrangian (\ref{L-QCD})
that the cutoff is much larger than $M$.  Integrating out the momentum scale
$M$ is equivalent to lowering the cutoff $\Lambda$ to a value lower than $M$.
We will argue that if $\Lambda$ is in the range
$Mv \ll \Lambda \ll M$, the resulting effective field theory can still be
described by a local lagrangian. In other words, the effects of
modes with momenta of order $M$ can be reproduced by local interactions
among the lower momentum modes.

Suppose the onium is in a virtual state that includes a quark with
relativistic momentum of order $M$.
Then that state is off its energy-shell
by an amount of order $M$, which is much larger than
the scale $Mv^2$ of the splittings between onium energy levels.
By the uncertainty principle, the lifetime of that highly virtual state
is less than or of order $1/M$.
In that short time, fields can propagate only
over distances of order $1/M$ that 
are pointlike on the scale $1/(Mv)$ of onium structure.
Thus the effects of virtual states that are excluded by a 
momentum cutoff in the range $Mv \ll \Lambda \ll M$
can be reproduced by
local interactions involving low-momentum modes.

The above argument applies equally well to virtual states that contain a
light parton with momentum of order $M$ in addition to the  
$Q \overline Q$ pair.  However,
it does not apply to virtual states obtained by the 
annihilation of the $Q \overline Q$ pair.  Such states can contain light 
partons with momenta of order $M$ without being far off the energy shell.
A momentum cutoff satisfying $\Lambda \ll M$ excludes these states,
but their effects cannot be reproduced in detail
by local interactions.  For example, the annihilation decay of
the onium produces light hadrons, some of which must have momenta of order 
$M$ and therefore parton constituents with momenta of order $M$.
With such modes excluded by the cutoff, we cannot
hope to describe the annihilation decays accurately.
Nevertheless, as we shall see in Section~\ref{sec:AnnDec},
the total annihilation  width of
an onium state can be described accurately.
Thus the momentum cutoff $\Lambda$ can be extended to light partons
at the expense of a restriction on the 
physical observables that can be described within the effective theory.

With a momentum cutoff $\Lambda$
that excludes relativistic $Q$ and $\overline Q$ states,
it is convenient to describe the heavy quark and antiquark by separate
2-component Pauli fields $\psi$ and $\chi$, rather than by a single
4-component Dirac field $\Psi$.
If we simply substitute $\Psi =  {\psi \choose \chi}$ into
the lagrangian (\ref{L-QCD}),
we obtain off-diagonal terms that couple $\psi^\dagger$ to $\chi$
and $\chi^\dagger$ to $\psi$,
allowing the creation and annihilation of $Q \overline Q$ pairs. 
We will argue that terms that change the numbers of heavy quarks
and antiquarks can be removed from the lagrangian and compensated by
terms that conserve the numbers of $Q$'s and $\overline Q$'s.
This is obvious for terms that create $Q \overline Q$ pairs, because 
a virtual state containing an additional $Q \overline Q$
must be off its  energy shell by an amount of order $M$.
As mentioned above, this is not completely true for terms that 
allow the $Q \overline Q$  pair in the onium to annihilate.  
The effects of states consisting of
gluons and light $q \bar q$ pairs that are produced by $Q \overline Q$
annihilation cannot be reproduced in detail
by local interactions.  However,
the effects of these states on sufficiently inclusive observables can be
described accurately.
Thus, with this restriction on physical observables, 
terms in the effective lagrangian that change the numbers of $Q$'s and  
$\overline Q$'s can be eliminated from the effective 
lagrangian.

For the Dirac term of the lagrangian (\ref{L-QCD}),
the decoupling of the fields $\psi$ and $\chi$ can be accomplished by a
unitary transformation called the
Foldy-Wouthuysen-Tani transformation.
For the case of a background gauge field,
it is straightforward to construct the transformation
that diagonalizes the lagrangian to any desired order in the
heavy quark velocity.  The simplest form of this transformation 
in the Dirac representation is
\begin{equation}
\Psi \;\to\;
\exp \left( -i \mbox{\boldmath $\gamma$} \cdot {\bf D}/2m_Q \right) \Psi.
\end{equation}
After this transformation, the heavy-quark term in the
lagrangian (\ref{L-QCD}) can be approximated by
\begin{equation}
{\psi \choose \chi}^\dagger
\left( \begin{array}{cc}
        -m_Q + iD_0 + {\bf D}^2/2m_Q & 0 \\
        0 & m_Q + iD_0 - {\bf D}^2/2m_Q
        \end{array} \right)
{\psi \choose \chi}.
\label{FWT}
\end{equation}
If we take ${\bf D}$ to scale like $Mv$,
the corrections to the entries in the matrix scaling like $M v^4$.

For the effective lagrangian with momentum cutoff $\Lambda$, 
the elimination of terms that change the numbers of $Q$'s and 
$\overline Q$'s is more
complicated than simply applying a unitary transformation. In addition to the
terms that are quadratic in $\psi$ and $\chi$,
the effective lagrangian also includes terms that are quartic and higher
in the heavy quark fields.  A further complication is that
gluon interactions
modify the coefficients of the terms produced by the 
Foldy-Wouthuysen-Tani transformation.
Nevertheless, by the general arguments presented above, one can
describe the low-energy $Q \overline Q$ sector of QCD by an 
effective field theory in which the numbers of heavy quarks and antiquarks 
are strictly conserved.

\subsection{Effective Field Theory}

One could in principle construct a nonrelativistic effective lagrangian that
describes the low-energy $Q \overline Q$ 
sector of QCD by starting with the lagrangian
(\ref{L-QCD}) and carrying out the sequence of two
transformations described in Section~\ref{sec:IntRel}.
The first is a renormalization group transformation
that removes modes with momenta greater than $\Lambda$.
The second is a transformation that removes interactions that change the 
numbers of heavy quarks and antiquarks.
Both of these steps would be extremely complicated to carry out
in practice.  Fortunately, there is an alternative to the
explicit construction of the effective lagrangian and that is to use the
strategy of ``effective field theory''.\cite{Georgi}
In this approach, the construction of the effective lagrangian
proceeds through the following steps:
\begin{enumerate}

\item Identify the fields that are required to describe the
low-energy excitations of the theory.

\item Identify the symmetries that one could maintain in the effective
theory by using a suitable cutoff and making appropriate field redefinitions.

\item Specify the accuracy to which low energy observables
in the original theory should be reproduced by the effective theory.

\item Write down the most general effective lagrangian that is
consistent with the symmetries, including all terms that are required to
reproduce the physics to the specified level of accuracy.

\item Determine the coefficients of those terms by matching low-energy
observables of the effective theory with those of the full theory.

\end{enumerate}

The effective field theory that is 
obtained by applying the above strategy to the low-energy
$Q \overline Q$ sector of QCD is called {\it nonrelativistic
QCD} (NRQCD).\cite{Caswell-Lepage}
The fields that are required to describe the low energy degrees of freedom are
the heavy quark and antiquark fields $\psi$ and $\chi$, the $SU(3)$
gauge fields $A_\mu$, and the Dirac fields for the light quarks.
The symmetries of NRQCD are the following:
\begin{list}{$\bullet$}{}

\item {\it $SU(3)$ gauge symmetry}.  This local symmetry requires
that the gluon fields enter into the effective lagrangian
only through the gauge-covariant derivatives 
$D_0$ and ${\bf D}$ and the QCD field strengths ${\bf E}$ and ${\bf B}$.

\item {\it rotational symmetry}.  A nonrelativistic description of the
heavy quark
necessarily breaks the Lorentz symmetry of QCD down to its rotational subgroup.

\item {\it charge conjugation} and {\it parity}.
These discrete symmetries of QCD are
also symmetries of the effective theory.  The charge conjugation
transformations of the heavy quark and antiquark fields are
\begin{equation}
\psi \;\to\; i \,(\chi^\dagger \sigma_2)^t, \qquad
\chi \;\to\; -i \,(\psi^\dagger \sigma_2)^t .
\end{equation}
The parity transformations are
\begin{equation}
\psi (t, {\bf r}) \;\to\; \psi (t, - {\bf r}), \qquad
\chi (t, {\bf r}) \;\to\; - \chi (t, - {\bf r}) .
\end{equation}

\item {\it heavy-quark phase symmetry}.  This symmetry guarantees the separate
conservation of the number of heavy quarks and antiquarks.  Its action on the
fields is
\begin{equation}
\psi \;\to\; e^{i \alpha} \psi, \qquad
\chi \;\to\; e^{i \beta} \chi .
\end{equation}

\end{list}

Having identified the symmetries of NRQCD, we can write down the most general
effective lagrangian that is consistent with these symmetries.  It has the
form
\begin{equation}
{\cal L}_{\rm NRQCD}
\;=\; {\cal L}_{\rm light}
\;+\; \psi^\dagger \left( iD_0 + {{\bf D}^2 \over 2M} \right) \psi
\;+\; \chi^\dagger \left( iD_0 - {{\bf D}^2 \over 2M} \right) \chi
\;+\; \delta {\cal L} ,
\label{L-NRQCD}
\end{equation}
where ${\cal L}_{\rm light}$ is the usual lagrangian that describes
gluons and light quarks.  The desired level of accuracy
is specified by the order in $v$ with which the onium  energy levels
must be reproduced by the effective theory.
The heavy quark terms that are shown explicitly in
(\ref{L-NRQCD}) are those that are
required to calculate the energy levels up to errors of order $Mv^4$.
The term $\delta{\cal L}$ in (\ref{L-NRQCD}) includes
the correction terms that must be added to decrease the errors
to order $M v^6$ or smaller.

The invariance of physical quantities under field redefinitions
can be exploited to eliminate some terms in the NRQCD lagrangian.
Gauge invariance requires that the gluon fields in $\delta {\cal L}$
appear only through the covariant derivatives  $D_0$ and ${\bf D}$.
However, a redefinition of the field $\psi$ can be used to eliminate terms
in which $D_0$ acts on $\psi$, and similarly for $\chi$.
Thus field redefinitions can be used to
eliminate all occurences of $D_0$ in $\delta {\cal L}$
except in the combination
$[D_0, {\bf D}] = i g {\bf E}$.  Because of these
field redefinitions, NRQCD will not
reproduce the low-energy behavior of the Green's functions of QCD.
It will only agree with full QCD for on-shell physical quantities.

The minimal form of NRQCD is obtained by setting $\delta {\cal L} = 0$ in
(\ref{L-NRQCD}).  It contains two parameters, the heavy-quark mass
parameter $M$ and the gauge coupling constant $g$.
These parameters can be tuned
as functions of the QCD coupling constant $\alpha_s$,
the heavy-quark mass parameter $m_Q$,
and the ultraviolet cutoff $\Lambda$ of NRQCD so that
the splittings between the onium energy
levels are reproduced up to errors of order $Mv^4$.  Since the 
energy splittings between radial excitations (such as $J/\psi$ and $\psi'$) 
and between orbital-angular-momentum excitations (such as $J/\psi$ and
$\chi_{cJ}$) scale like $Mv^2$, 
these are reproduced up to errors of relative order
$v^2$.  Spin splittings in heavy quarkonium, such as the splitting between
the lowest $^1S_0$ and $^3S_1$ states ($\eta_c$ and $J/\psi$ for charmonium),
scale like $Mv^4$.  These splittings vanish in minimal NRQCD
due to the following symmetry:

\begin{list}{$\bullet$}{}

\item {\it heavy-quark spin symmetry}.  Under this symmetry,
the two spin components of the heavy quark and the two spin components of the
antiquark are mixed by independent unitary transformations:
\begin{equation}
\psi \;\to\; U \psi, \qquad
\chi \;\to\; V \chi,
\end{equation}
where $U$ and $V$ are SU(2) matrices.  This is only
an approximate symmetry of the complete NRQCD lagrangian, holding
up to corrections of relative order $v^2$.
\end{list}

If we wish to reduce  the errors in the quarkonium energy levels
to smaller than order $Mv^4$, it is necessary to add additional
terms $\delta {\cal L}$ to the lagrangian in (\ref{L-NRQCD}).
Using the velocity-scaling rules that are discussed in
Section~\ref{sec:vscaling}, it can be shown that the terms that
are necessary and sufficient to reduce the errors to order $M v^6$ are
\begin{eqnarray}
\delta{\cal L}
&=& {c_1 \over 8M^3} \psi^\dagger ({\bf D}^2)^2 \psi
\;+\; {c_2 \over 8M^2}
\psi^\dagger ({\bf D} \cdot g {\bf E} - g {\bf E} \cdot {\bf D}) \psi
\nonumber \\
&& \;+\; {c_3 \over 8M^2}
\psi^\dagger (i {\bf D} \times g {\bf E}
        - g {\bf E} \times i {\bf D}) \cdot \mbox{\boldmath $\sigma$} \psi
\;+\; {c_4 \over 2M}
\psi^\dagger (g {\bf B} \cdot \mbox{\boldmath $\sigma$}) \psi
\nonumber \\
&& \;+\: {\rm \; charge \; conjugate \; terms} ,
\label{deltaL}
\end{eqnarray}
where $c_1$, $c_2$, $c_3$, and $c_4$ are dimensionless coefficients.
We will refer to the terms in (\ref{deltaL}) 
as the {\it $v^2$-improvement terms} in the NRQCD lagrangian.
The two terms in (\ref{deltaL}) that contain the Pauli
matrix $\mbox{\boldmath{$\sigma$}}$ break the spin symmetry
of minimal NRQCD.  They give spin splittings that scale like $Mv^4$ and
are accurate up to errors of relative order $v^2$.
Splittings between radial excitations  and splittings
between orbital-angular-momentum excitations are reproduced up to 
errors of relative order $v^4$.

\subsection{Velocity-scaling rules}
\label{sec:vscaling}

The relative importance of the terms in the NRQCD lagrangian can be deduced
from the self-consistency of the quantum field equations for minimal NRQCD
and from the basic qualitative features of quarkonium.\cite {L-M-N-M-H}  
The results of this analysis are summarized  by the velocity-scaling rules 
in Table~\ref{tab:vscaling}.
The magnitude of a matrix element of a local gauge-invariant operator
between quarkonium states can be estimated by multiplying
the appropriate factors from Table~\ref{tab:vscaling}.  The scaling
with $M$ follows simply from dimensional analysis, so the nontrivial content
of Table~\ref{tab:vscaling} is the scaling with $v$.

The first few lines in Table~\ref{tab:vscaling} can be derived very
easily.  The expectation value of the number operator
$\int d^3x \; \psi^\dagger \psi$ in a quarkonium state 
$| H \rangle$ is very close to 1:
\begin{equation}
\langle H| \int d^3 x \; \psi^\dagger \psi \; | H\rangle \;\approx\; 1.
\end{equation}
We have normalized the quarkonium state so that
$\langle H | H\rangle = 1$. From
the fact that a quarkonium state can be localized to within a region
$1/(Mv)^3$, we conclude that $\psi$ must scale like $(Mv)^{3/2}$.  The
expectation value of the kinetic energy term in the NRQCD hamiltonian
scales like $Mv^2$:
\begin{equation}
\langle H| \int d^3x \; \psi^\dagger ({\bf D}^2/2M) \psi \; |H\rangle
\;\sim\; M v^2.
\end{equation}
This implies that ${\bf D}$ must scale like $Mv$.  
The fact that $D_0$ scales like $Mv^2$  when acting on $\psi$
then follows immediately from the field equation for $\psi$:
\begin{equation}
\left( iD_0 - {{\bf D}^2 \over 2M}\right) \psi \;=\; 0.
\end{equation}

\begin{table}[t]
\caption{Estimates of the magnitudes of NRQCD operators for matrix elements
        between heavy-quarkonium states.
        \label{tab:vscaling}}
\vspace{0.2cm}
\begin{center}
\begin{tabular}{|l|l|}
\hline
 Operator                               & Estimate      \\
\hline
 $\psi$                                 &  $(Mv)^{3/2}$ \\
 $\chi$                                 &  $(Mv)^{3/2}$ \\
 $D_0$     (acting on $\psi$ or $\chi$) &  $M v^2$      \\
 ${\bf D}$                              &  $M v$        \\
 $g {\bf E}$                            &  $M^2v^3$     \\
 $g {\bf B}$                            &  $M^2v^4$     \\
 $g A_0$      (in Coulomb gauge)        &  $M v^2$      \\
 $g{\bf A}$ \ (in Coulomb gauge)        &  $M v^3$      \\
\hline
\end{tabular}
\end{center}
\end{table}

The estimates for $gA_0$ and $g{\bf A}$ in Table \ref{tab:vscaling} are specific
to Coulomb gauge, which is defined by
$\mbox{\boldmath $\nabla$} \cdot {\bf A} = 0$.
As shown below, the field equations in this gauge indicate
that the effects of the vector potential ${\bf A}$ are suppressed by a factor
of $v$ relative to the scalar potential $A_0$.  The dominant terms in the
field equations for $\psi$ and $A_0$ are therefore
\begin{eqnarray}
\left( i \partial_0 - gA_0
        + {\mbox{\boldmath $\nabla$}^2 \over 2M} \right) \psi
& \approx & 0 ,
\label{feq-psi}
\\
\mbox{\boldmath $\nabla$}^2 g A_0 + g^2 \psi^\dagger \psi & \approx & 0.
\label{feq-A0}
\end{eqnarray}
In (\ref{feq-psi}), the balance between the kinetic
energy and the potential energy
represented by the $A_0$ term requires that $gA_0$ scale like
$Mv^2$.  On the other hand, assuming that a gradient acting on $A_0$ scales
like $Mv$, (\ref{feq-A0}) requires that $gA_0$ scale like $g^2Mv$.
These two estimates are consistent if the effective coupling constant
$\alpha_s = g^2/4\pi$ at the scale $Mv$ scales like $v$.  This
is identical to the naive estimate (\ref{v-estimate})
that followed from balancing the kinetic energy and the Coulomb term 
in the potential energy.  Since this scaling relation follows
simply from the consistency of the field equations, it applies
to charmonium and bottomonium even though 
perturbation theory at the scale $Mv$ is of questionable validity. 
The neglect of terms involving ${\bf A}$
in the field equations (\ref{feq-psi}) and (\ref{feq-A0}) is justified by the
field equation for ${\bf A}$, for which the dominant terms are
\begin{equation}
\left ( \partial_0^2 - \mbox{\boldmath $\nabla$}^2 \right) g{\bf A}
\;-\; g A_0 \mbox{\boldmath $\nabla$} gA_0
\;-\; { g^2 \over M} \psi^\dagger \mbox{\boldmath $\nabla$} \psi
\;\approx\; 0.
\label{feq-A}
\end{equation}
The last two terms in (\ref{feq-A}) scale like $M^3 v^5$ and $g^2 M v^4$,
respectively, and they are comparable if $g^2$ scales like $v$.
Assuming that a gradient acting on ${\bf A}$ scales like $Mv$,
we obtain the estimate $M v^3$ for $g{\bf A}$ in Table~\ref{tab:vscaling}.

Using the estimates for $gA_0$ and $g{\bf A}$ in Coulomb gauge, we can obtain
estimates for the field strengths.  In Coulomb gauge, the dominant term in the
chromoelectric field strength ${\bf E}$ is $-\mbox{\boldmath $\nabla$}
A_0$, and  the
resulting estimate for $g {\bf E}$ is $M^2v^3$.  The dominant term in the
chromomagnetic field strength ${\bf B}$ is
$\mbox{\boldmath $\nabla$} \times {\bf A}$, which leads to
the estimate $M^2v^4$ for $g{\bf B}$.  These estimates for $g{\bf E}$
and $g {\bf B}$, although derived in Coulomb gauge, hold in general
for matrix elements of gauge-invariant operators.

According to the velocity-scaling rules in Table~\ref{tab:vscaling},
the terms in the lagrangian density in (\ref{L-NRQCD}) for
minimal NRQCD scale like $M^4 v^5$.  Multiplying by a volume factor of
$1/(M v)^3$, we find that quarkonium energies scale like $Mv^2$.
Each of the terms in $\delta{\cal L}$ given in (\ref{deltaL})
scales like $M^4 v^7$ and therefore contributes to onium energies
at order $Mv^4$.  All other terms that can be added to
the NRQCD lagrangian give contributions of order $M v^6$ or smaller.

The validity of the  velocity-scaling relations has been demonstrated 
convincingly by nonperturbative calculations of the bottomonium and 
charmonium spectrum using Monte Carlo simulations of 
lattice NRQCD.\cite{NRQCD:spectrum}
The two parameters of minimal NRQCD can be tuned to give the spin-averaged
spectrum to an accuracy of about 30\% for charmonium and about 10\% for 
bottomonium.  When the $v^2$-improvement terms are included, the 
errors decrease to about 10\% for charmonium and to about 1\%
for bottomonium. These terms also give spin splittings that
are accurate to about 30\% for charmonium and to about 
10\% for bottomonium.

\subsection{Fock state expansion}
\label{sec:Fock}

The simplest intuitive picture of quarkonium is 
that it is a bound state consisting of a
$Q$ and $\overline Q$ with very little probability of containing additional
gluons or $q \bar q$ pairs.  This simple picture is in fact
realized in Coulomb gauge.  Using the velocity-scaling rules of
section~\ref{sec:vscaling},
one can quantify the probabilities of Fock states containing additional gluons
by determining how they scale with $v$.

The Coulomb gauge ($\mbox{\boldmath $\nabla$} \cdot {\bf A} = 0$)
is a physical gauge with no negative norm states, 
a necessary condition for a sensible Fock space.
In this gauge, the scalar potential $A_0$ does not propagate.
Dynamical gluons are created
and destroyed by the vector potential ${\bf A}$.
In Coulomb gauge, the lagrangian (\ref{L-NRQCD}) can be reorganized
as an expansion in powers of $v$.
The powers of $v$ can be made explicit\cite{Luke-Manohar}  by rescaling
the space-time coordinates ${\bf r}$ and $t$
by $1/(Mv)$ and $1/(Mv^2)$, respectively,
rescaling the fields $\psi$ and $\chi$ by $(Mv)^{3/2}$,
and rescaling the fields $A_0$ and ${\bf A}$ by 
$Mv^{3/2}$.
The terms in the NRQCD lagrangian that are of order $v^0$
after such a rescaling are
\begin{equation}
{\cal L}_0
\;=\; {\cal L}_{\rm light}
\;+\; \psi^\dagger \left( i \partial_0 - g A_0
        + { \mbox{\boldmath $\nabla$}^2 \over 2M } \right) \psi
\;+\; \chi^\dagger \left( i \partial_0 - g A_0
        - { \mbox{\boldmath $\nabla$}^2 \over 2M } \right) \chi .
\label{LC-0}
\end{equation}
This lagrangian in Coulomb gauge
can be used to calculate quarkonium energy levels to the same
accuracy as the gauge-invariant lagrangian of minimal NRQCD.
The terms in the NRQCD lagrangian that are of order $v$ 
after the rescaling are
\begin{eqnarray}
{\cal L}_1
&=& - {1 \over M}
\psi^\dagger (i g {\bf A} \cdot \mbox{\boldmath $\nabla$} ) \psi
\;+\; {c_4 \over 2M}
\psi^\dagger (\mbox{\boldmath $\nabla$} \times g {\bf A})
        \cdot \mbox{\boldmath $\sigma$} \psi
\nonumber \\
&& \;+\; {\rm \; charge \; conjugate \; terms} .
\label{LC-1}
\end{eqnarray}
At first order in perturbation theory, these terms give transitions from
$Q \overline Q$ Fock states to states that contain a dynamical gluon.
The expectation values of the terms in (\ref{LC-1})
between $| Q \overline Q \rangle$ Fock states
vanish, so ${\cal L}_1$ first contributes to quarkonium energy levels
at second order in perturbation theory, giving shifts of order $Mv^4$.
The terms in the Coulomb-gauge lagrangian that are of order $v^2$ 
after rescaling are
\begin{eqnarray}
{\cal L}_2
&=& - {1 \over 2M}
\psi^\dagger (g {\bf A})^2 \psi
\;+\; {c_1 \over 8M^3} \psi^\dagger (\mbox{\boldmath $\nabla$}^2)^2 \psi
\nonumber \\
&& \;+\; {c_2 \over 8M^2}
\psi^\dagger ( - \mbox{\boldmath $\nabla$}^2 g A_0 ) \psi
\;-\; {c_3 \over 4M^2}
\psi^\dagger (\mbox{\boldmath $\nabla$} g A_0 ) \times
        \mbox{\boldmath $\nabla$} \cdot \mbox{\boldmath $\sigma$} \psi
\nonumber \\
&& \;+\; {c_4 \over 2M}
\psi^\dagger (i g {\bf A} \times g {\bf A})
        \cdot \mbox{\boldmath $\sigma$} \psi
\;+\: {\rm \; charge \; conjugate \; terms} .
\label{LC-2}
\end{eqnarray}
The first term comes from expanding out the covariant derivative ${\bf D}$ in
(\ref{L-NRQCD}), while the last four terms in (\ref{LC-2}) come from 
the $v^2$-improvement terms in (\ref{deltaL}). Energy levels
calculated in Coulomb gauge using the lagrangian
${\cal L}_0 + {\cal L}_1 + {\cal L}_2$
will differ only at order $Mv^6$ from the energy levels calculated using the
gauge-invariant NRQCD lagrangian (\ref{L-NRQCD}) with $\delta {\cal L}$ given
by (\ref{deltaL}).

We now consider the Fock state expansion in
Coulomb gauge for a quarkonium state $|H\rangle$.  It has the schematic form
\begin{equation}
| H \rangle \;=\;  \psi^H_{Q \overline Q} |Q \overline Q \rangle 
	\;+\; \psi^H_{Q \overline Qg} |Q \overline Qg\rangle 
	\;+\; \dots ,
\label{Fock}
\end{equation}
where spin and color indices and momentum arguments have all been suppressed.
The dominant Fock state $|Q\overline Q\rangle$ consists of a
$Q$ and $\overline Q$ in a color-singlet state with definite angular momentum
quantum numbers $^{2S + 1}L_J$.  The higher Fock states, such as
$| Q \overline Q g \rangle $, include dynamical gluons or light $q \bar q$ 
pairs.  Since the lagrangian
${\cal L}_0$ in (\ref{LC-0}) does not include any terms that couple ${\bf A}$
to $\psi$ or $\chi$, the probabilities for higher Fock states are suppressed
by powers of $v$.

The $|Q \overline Q g\rangle$ states with the highest probabilities are those
that couple to the dominant $|Q \overline Q \rangle$ state via the lagrangian
${\cal L}_1$ in (\ref{LC-1}).  We first consider the term
$\psi^\dagger (i g {\bf A} \cdot \mbox{\boldmath $\nabla$}) \psi$.
We refer to a transition that proceeds
via this term  or its charge conjugate as an {\it electric transition}.
An electric transition from the dominant $|Q \overline Q \rangle$ Fock state
produces $|Q \overline Q g \rangle$ states for which the
angular-momentum quantum numbers of
the $Q \overline Q$ pair satisfy the selection rules $\Delta L = \pm 1$ and
$\Delta S = 0$.  The simplest way to determine the probabilities 
of these Fock states is to use the fact that
a second-order perturbation in ${\cal L}_1$ changes the mass of the
onium state by an amount of order $Mv^4$.  This mass shift can be expressed
as the product of the energy $E$ of the virtual $|Q \overline
Qg\rangle$ state multiplied by its probability $P$.  If the energy of the
dynamical gluon is of order $Mv$, then $E \sim Mv$ and we
find that $P \sim v^3$. If the gluon has energy of order $Mv^2$ or less,
then $E \sim Mv^2$ and we obtain $P \sim v^2$.
We conclude that $|Q \overline Qg\rangle$
states which satisfy the selection rules $\Delta L = \pm 1$ and $\Delta S = 0$
are dominated by very soft dynamical gluons with momenta of order $Mv^2$ or
less and have probabilities of order $v^2$.

We next consider the term
$\psi^\dagger(\mbox{\boldmath $\nabla$} \times g {\bf A})
        \cdot \mbox{\boldmath $\sigma$} \psi$ in (\ref{LC-1}).
We refer to a transition that proceeds
via this term or its charge conjugate as a {\it magnetic transition}.
A magnetic transition from the dominant
$| Q \overline Q \rangle$ Fock state produces
$|Q \overline Q g \rangle$ states that satisfy
the selection rules $\Delta L = 0$ and $\Delta S = \pm 1$.
We can use the same argument as before to determine the probabilities
of these Fock states, except that we must take into account the fact
that the transition amplitude from the term
$\psi^\dagger (\mbox{\boldmath $\nabla$} \times g{\bf A})
        \cdot \mbox{\boldmath $\sigma$} \psi$
is weighted by the momentum of the gluon.  If the
gluon has energy of order $Mv$, the mass shift from
a second-order perturbation in ${\cal L}_1$ is given correctly by the
velocity-scaling rules to be of order $Mv^4$.  Since 
the virtual $|Q \overline Q g \rangle$ state has energy $E \sim Mv$,
we obtain a probability $P \sim v^3$.  The contribution to the 
mass shift from a gluon with energy of order
$Mv^2$ is suppressed by a factor of $v^2$
from the transition amplitudes and is therefore of order $Mv^6$.
Taking $E \sim Mv^2$,
we obtain $P \sim v^4$.  We conclude that $|Q \overline Qg\rangle$ states
which satisfy the selection rules $\Delta L = 0$ and $\Delta S = \pm 1$ are
dominated by dynamical gluons with momenta of order $Mv$ and have
probabilities of order $v^3$.

Similar arguments can be used to determine the magnitudes of the
probabilities for other Fock states.  Any such state can be reached by a 
sequence of electric transitions and  zero or one magnetic transition.
Electric transitions obey the selection rules
$\Delta L = \pm 1$ and $\Delta S = 0$, while magnetic transitions
satisfy  $\Delta L = 0$ and $\Delta S = \pm 1$. Both electric and magnetic
transitions change the color state of a
color-singlet $ Q \overline Q$ pair to color-octet,
and they change  the color state of a
color-octet $Q \overline Q$ pair to either color-singlet
or color-octet.  The probability of a particular Fock state is determined by
the color and angular-momentum quantum numbers of the
$Q \overline Q$ pair in that state.  If that
Fock state can be reached
from the dominant $| Q \overline Q \rangle$ Fock state
by a sequence of $E$ electric transitions, then its probability
scales like $v^{2E}$.  If it can be reached
by a sequence of $E$ electric transitions and a magnetic transition,
then its probability scales like $v^{2E+3}$.

\subsection{Matching of NRQCD and QCD}

The NRQCD lagrangian contains adjustable parameters that must be tuned in
order that its predictions for low-energy observables in the 
$ Q \overline Q$ sector
agree with those of QCD.  In the minimal NRQCD lagrangian,
there are two parameters: $g$ and $M$.
In addition to these parameters, the definition of NRQCD
requires an ultraviolet cutoff $\Lambda$
to remove ultraviolet divergences.
In the $v^2$-improved lagrangian obtained by adding the terms in
(\ref{deltaL}), there are 6 parameters,  $g$, $M$, $c_1$,
$c_2$, $c_3$, and $c_4$, in addition to the ultraviolet cutoff.
The determination of the parameters in the NRQCD lagrangian is called
matching.

One could in principle determine the $N$ parameters in the NRQCD lagrangian 
by tuning them so that the  masses of $N$ states in NRQCD 
match the corresponding masses
in full QCD.  For example, the parameters $g$ and $M$ of
minimal NRQCD could be determined by matching the mass splittings 
between $J/\psi$ and $\psi'$ and between $J/\psi$ and $\chi_{cJ}$.
Since the masses are sensitive to long-distance effects,
they must be calculated nonperturbatively.
The only reliable nonperturbative method
that is currently available is Monte Carlo simulations of lattice NRQCD.  
While the masses in full QCD could in principle be computed nonperturbatively
using lattice simulations, it is easier to 
take them directly from experiment. 
Using masses to tune the NRQCD parameters is an example of 
{\it nonperturbative matching}.

Nonperturbative matching would become increasingly
difficult as we strive for higher accuracy by adding more improvement terms.
The determination of the parameters in the $v^2$-improved lagrangian 
would require the nonperturbative calculation of 6 masses 
as functions of 6
independent parameters.  Fortunately, the asymptotic freedom of QCD provides
an alternative, and that is {\it perturbative matching}.
This matching procedure is based on the fact that QCD and NRQCD are 
equivalent except on distance scales of order $1/M$ 
where perturbative QCD is by assumption  accurate.
The procedure for perturbative matching is the following:
\begin{enumerate}

\item Use perturbative QCD to calculate scattering amplitudes
between asymptotic $Q$, $\overline Q$, and gluon states with momenta ${\bf k}$
much less than $M$ as functions of $\alpha_s$ and $m_Q$
and expand them in powers of ${\bf k}/m_Q$.

\item  Use perturbative NRQCD to calculate the same scattering amplitudes
in terms of the parameters in the NRQCD lagrangian
and expand them in powers of ${\bf k}/M$.

\item Adjust the NRQCD parameters so that the scattering amplitudes match
to the desired order in ${\bf k}/m_Q$, which we take to be of order $v$.

\end{enumerate}
It is essential to match scattering amplitudes or other
physical observables rather than Green
functions, because the construction of NRQCD involves field redefinitions.
Such redefinitions can change the off-shell Green functions of the theory,
but they leave on-shell physical observables unchanged.

In present calculations in lattice  NRQCD, the parameters 
are determined by a combination of nonperturbative and
perturbative matching.  The coefficients $c_1$, $c_2$, $c_3$, and $c_4$
of the $v^2$-improvement terms are generally determined by
perturbative matching, while the parameters $g$ and $M$ are determined by
the nonperturbative matching of masses in the onium spectrum.
Perturbative matching calculations can be used to relate these parameters 
to the fundamental parameters of QCD. 
By combining these perturbative matching relations with lattice NRQCD 
calculations of the bottomonium spectrum, the 
QCD coupling constant $\alpha_s$ and the bottom quark mass $m_b$ 
have been determined with high precision.\cite{NRQCD:alphas}

The method of perturbative matching is somewhat paradoxical.  We have assumed
that $M$ is large enough that perturbation theory is accurate at the scale
$M$.  We allow for the scale $Mv$ to be small enough that
perturbation theory is not reliable at that scale.  If that is the case,
perturbative calculations
in NRQCD would never give accurate results for physical observables,
since NRQCD only reproduces full
QCD accurately at scales of order $Mv$ or less. 
Nevertheless, a comparison of perturbative
calculations in NRQCD and full QCD can be used to accurately determine the
parameters in the NRQCD lagrangian.  The reason for this is
that the tuning of the parameters of NRQCD that makes
this theory equivalent to QCD
at momenta of order $Mv$ or smaller also makes
the perturbative approximations to these theories equivalent.
Perturbation theory  breaks down in precisely the
same way for both theories, predicting
among other things, the existence of asymptotic states consisting of
isolated quarks and gluons. Since the parameters in the NRQCD
lagrangian are sensitive only to momenta on the order of $M$ where
perturbative QCD is accurate, they can be correctly
determined by matching perturbative calculations in QCD and NRQCD.

As an illustration of perturbative matching, we consider the simplest
perturbative observable.  This is the energy-momentum relation for the heavy
quark, which is given by the location of the pole in the heavy-quark
propagator.  At tree level in full QCD, the energy-momentum relation is
\begin{equation}
E \;=\; \sqrt {m^2_Q + p^2}
\;=\; m_Q \;+\; {p^2 \over 2m_Q} \;-\; {p^4 \over 8 m^3_Q}
\;+\; \dots .
\label{E-QCD}
\end{equation}
At tree level in NRQCD, we can read off the energy momentum relation from the
lagrangian (\ref{L-NRQCD}):
\begin{equation}
E \;=\; {p^2 \over 2M} \;-\; c_1 {p^4 \over 8 M^3} \;+\; \dots .
\label{E-NRQCD}
\end{equation}
By matching the expressions (\ref{E-QCD}) and (\ref{E-NRQCD}) we find
\begin{equation}
M \;=\; m_Q, \qquad c_1\;=\; 1.
\label{M-cl}
\end{equation}
If the energy-momentum relations are computed to higher order in perturbation
theory, the matching will give perturbative corrections to the results in
(\ref{M-cl}).  Since the parameters $M$ and $c_1$ are sensitive only to short
distances of order $1/m_Q$ or smaller, the corrections can be expressed as
power series in $\alpha_s(m_Q)$.

\section{Annihilation Decays of Heavy Quarkonium}
\label{sec:AnnDec}

\subsection{Decay of $\eta_c$ in the Color-Singlet Model}

A simple intuitive picture of the annihilation decay of a quarkonium state
is that it proceeds through the annihilation of the $Q \overline Q$
pair in the dominant Fock state into gluons and light $q \bar q$ pairs.
These light partons ultimately hadronize into the observed final states
that consist of light hadrons.  The inclusive annihilation rate
of the $Q \overline Q$  pair can be plausibly calculated using
perturbative QCD.  By combining that perturbative calculation
with a phenomenological wavefunction for the  dominant
$| Q \overline Q \rangle$ Fock  state,
we can calculate the annihilation decay rate of the quarkonium.
This model for annihilation decays is called the {\it color-singlet model}.

The simplest illustration of the color-singlet model is the
calculation of the decay rate of the $\eta_c$.
In the color-singlet model, the $\eta_c$ is modeled by
a $c \bar c$ pair in a color-singlet
$^1S_0$ state.  Its wavefunction is the product of a color factor
$\delta_{ij}/\sqrt 3$, a spin factor
$(\uparrow \downarrow - \downarrow \uparrow) / \sqrt 2$,
and a coordinate-space wavefunction $\psi({\bf r})= R(r)/\sqrt{4 \pi}$.
A $c \bar c$ pair in such a state can annihilate into two gluons.
If we assume that the two gluons hadronize into light hadrons with probability
1, the decay rate can be written as
\begin{equation}
\Gamma (\eta_c) \;=\;
{1 \over 2M_{\eta_c}} \int {d^3k \over (2{\pi})^3 2k} \,
{2 \pi \delta (M_{\eta_c} - 2|{\bf k}|) \over  M_{\eta_c}} \,
\big| {\cal T} [ \eta_c \to g ({\bf k}) g (- {\bf k}) ] \big|^2.
\label{Gam-eta:0}
\end{equation}
The T-matrix element for this decay can be expressed in terms of the
momentum-space wavefunction $\psi({\bf q})$ of the $\eta_c$:
\begin{equation}
{\cal T} [ \eta_c \to g ({\bf k}) g (- {\bf k}) ] \;=\;
{1 \over \sqrt{2 M_{\eta_c}}} \int {d^3 q \over (2 \pi)^3} \;
\psi ({\bf q}) \; {\cal T} [ c({\bf q}) \bar c(-{\bf q})
        \to g({\bf k}) g(-{\bf k}) ],
\label{T-eta}
\end{equation}
where we have suppressed all color and spin indices.
The $c \bar c$ 
annihilation amplitude ${\cal T}$ varies significantly with ${\bf q}$
only when $|{\bf q}|$ is on the order of $m_c$ or larger.
The wavefunction has significant support only for $|{\bf q}|$ of order
$m_c v$.  Since ${\cal T}$ 
is almost independent of ${\bf q}$ for such small values of  $|{\bf q}|$,
we can set ${\bf q} = 0$ in the annihilation amplitude.
The resulting expression for the decay rate has a factored form:
\begin{eqnarray}
\Gamma(\eta_c) &\approx&
{1 \over (2M_{\eta_c})^2}
\left| \int {d^3q \over (2 \pi)^3} 
	\psi({\bf q}) \right|^2
\nonumber \\
&& \hspace{-.2in} \times
\int{d^3k \over (2 \pi)^3 2 k} \; 
{ 2 \pi \delta (M_{\eta_c} - 2 | {\bf k}|) \over M_{\eta_c} }\;
\big| {\cal T} [ c(0) \bar c(0) \to g({\bf k}) g(-{\bf k}) ] \big|^2.
\label{Gam-eta:2}
\end{eqnarray}
The integral over ${\bf q}$ gives the wavefunction 
evaluated at the origin, $R(0)/\sqrt{4 \pi}$.
The integral over ${\bf k}$ 
in (\ref{Gam-eta:2}) can
be calculated from the lowest order QCD Feynman diagrams for $c \bar c
\to g g$.  The final expression for the decay rate is
\begin{equation}
\Gamma (\eta_c) \;\approx\;
{8 \alpha_s^2 \over 3M_{\eta_c}^2} |R(0)|^2 .
\label{Gam-eta:3}
\end{equation}

\subsection{Decay of $\eta_c$ in NRQCD}

The decay rate of the $\eta_c$ into individual final states
consisting of light hadrons can
not be described within the framework of NRQCD.
One obstacle is that we have imposed a symmetry on the
effective field theory that guarantees the separate conservation 
of the numbers of $c$'s and $\bar c$'s and therefore forbids 
the annihilation process $c \bar c \to g g$.
Furthermore, in the construction of NRQCD, we have integrated out gluons
with momenta on the order of $m_c$.
Even if we relax the definition of the effective theory
to allow gluons with momenta of order $m_c$ and interaction
terms in the lagrangian that allow $c \bar c$ annihilation, 
we cannot described annihilation decays accurately, 
because the interaction of a 
$c$ or $\bar c$ with a gluon of momentum $m_c$ cannot be described 
accurately in a local nonrelativistic theory.

While the decay rate of the $\eta_c$ into a specific final state consisting of
light hadrons cannot be described within NRQCD, the total decay rate can.
The conservation of the number of $c$ and $\bar c$ is not an obstacle,
because the optical theorem can be used to express the inclusive annihilation
rate in terms of an amplitude that conserves the numbers of $c$ and
$\bar c$.  As an illustration, applying the optical theorem to the
expression (\ref{Gam-eta:0}), we find that the decay rate can be written
\begin{eqnarray}
\Gamma(\eta_c) &=& {1 \over (2 M_{\eta_c})^2}
\int {d^3q \over (2 \pi)^3} \int {d^3 q' \over (2 \pi)^3} \;
\nonumber \\
&& \hspace{.5in} \times
\psi({\bf q}) \;
2 \, {\rm Im} {\cal T} [ c({\bf q}) \bar c(-{\bf q})
        \to c({\bf q}') \bar c(-{\bf q}') ] \; \psi^*({\bf q}').
\label{Gam-eta:4}
\end{eqnarray}
The process $c \bar c \to c \bar c$ conserves the numbers of $c$'s
and $\bar c$'s.  The factored form (\ref{Gam-eta:2})
is recovered by using the
fact that the imaginary part of the T-matrix element for 
$c\bar c \to c \bar c$ 
is insensitive to momenta ${\bf q}$ and ${\bf q}'$ on the
order of $Mv$ where the wavefunctions have their support:
\begin{equation}
\Gamma (\eta_c)
\;\approx\; {1 \over (2 M_{\eta_c})^2}
\left| \int {d^3q \over (2 \pi)^3} \psi({\bf q}) \right|^2 \,
2 \, {\rm Im}{\cal T} [c(0) \bar c(0) \to c(0) \bar c(0)].
\label{Gam-eta:5}
\end{equation}

The integral over ${\bf q}$ in (\ref{Gam-eta:5}) 
gives the square of the wavefunction 
evaluated at the origin.  This reflects the fact that the initial
$c$ and $\bar c$
must have spacetime separations of order $1/M$ in order to annihilate,
and this separation
is small compared to the length scale $1/(Mv)$ of the
wavefunction.  That the $c \bar c$ pair 
must have a spacetime separation of order $1/M$ 
follows from the fact that the Feynman diagrams for the annihilation
process involve a heavy quark propagator that is off its mass-shell 
by an amount of order $M$.
The same argument implies that the final $c$ and $\bar c$ in the 
T-matrix element in (\ref{Gam-eta:5}) 
must have a space-time separation of order $1/M$.
While it is not quite as obvious, the space-time separation of the
annihilation points for the 
initial $c \bar c$ pair and the final $c \bar c$ pair must also
be of order $1/M$.  This follows from the requirement that the 
wavefunctions of the annihilation gluons must overlap, which
localizes the production point of a gluon
to within its wavelength $1/M$.

The fact that $c \bar c$ annihilation occurs within a region whose size is of
order $1/M$ provides a clue as to how the effects of annihilation can be taken
into account in NRQCD.  All modes with momenta of order $M$ that can be
sensitive to the length scale $1/M$ have been removed from this effective
theory.  Thus the effects of the annihilation can be reproduced by including
in the NRQCD lagrangian (\ref{L-NRQCD}) a local 4-fermion interaction term
that destroys a $c \bar c$ pair and creates it again.  The specific term
that is relevant to $\eta_c$ decay is
\begin{equation}
\delta{\cal L}
\;=\; {f \over M^2} \psi^\dagger \chi \chi^\dagger \psi ,
\label{L-1S0}
\end{equation}
where the coefficient $f$ is dimensionless.
The term (\ref{L-1S0}) annihilates a $c \bar c$ pair
in a color-singlet $^1S_0$ state and then creates a $c \bar c$ pair
in the same state.

The dimensionless coefficient $f$ in (\ref{L-1S0}) can be determined by
perturbative matching of the $c \bar c \to c \bar c$ scattering
amplitudes in full QCD and NRQCD.  In full QCD, the scattering amplitude
includes box diagrams of order $\alpha_s^2$ 
in which the scattering proceeds through intermediate states consisting 
of two gluons.
In NRQCD,
this contribution to the scattering amplitude can only be reproduced by
4-fermion interactions such as those in (\ref{L-1S0}).  By matching the
NRQCD scattering amplitude from the term (\ref{L-1S0}) with the
annihilation part of the scattering amplitude in full QCD,
we can determine the coefficient $f$.  Since the $c$
and $\bar c$ can annihilate into two on-shell gluons, the QCD 
scattering amplitude has an imaginary part.  The coefficient
$f$ in (\ref{L-1S0}) must therefore have an imaginary part and it is
particularly simple to calculate.  The result is
\begin{equation}
{\rm Im} \, f \;=\; {2\pi \alpha_s^2(m_c) \over 9} .
\label{Imf}
\end{equation}
Since this coefficient is sensitive only to distances of order
$1/m_c$ or smaller, the running coupling constant is evaluated at the scale
$m_c$.  The fact that coefficients in the NRQCD lagrangian have imaginary parts
implies that the hamiltonian for NRQCD is not hermitian.  This is perfectly
natural, since we have removed states from the theory that are essential
for exact unitarity.  In particular, we have eliminated the light partons
with momenta on the order
of $m_c$ that can be produced by the annihilation of the $c$ and $\bar c$.

We now consider the effect of the correction terms (\ref{L-1S0}) on the energy
of the $\eta_c$.  If that term is treated as a first-order perturbation, the
resulting correction to the energy of the $\eta_c$ is
\begin{equation}
\Delta E_{\eta_c}
\;=\; -{f \over M^2}
{ \langle \eta_c | \psi^\dagger \chi \chi^\dagger \psi|\eta_c \rangle
        \over 2 M_{\eta_c}},
\end{equation}
where we have assumed that the state $|\eta_c \rangle$ has the standard
relativistic normalization.
Since the coefficient $f$ has an imaginary part, this energy shift has an
imaginary part.  A state whose energy has a small imaginary part $-\Gamma/2$
should be interpreted as  a resonance of width $\Gamma$.  Thus the
width of the $\eta_c$ due to the term (\ref{L-1S0}) is
\begin{equation}
\Gamma(\eta_c)
\;=\; {1 \over 2 M_{\eta_c}}
{4 \pi \alpha_s^2(m_c) \over 9 m_c^2} \langle \eta_c | \psi^\dagger \chi
\chi^\dagger \psi|\eta_c\rangle
\label{Gam-eta}.
\end{equation}
The connection with the result (\ref{Gam-eta:4}) from the color-singlet
model is made
by inserting a complete set of states
between $\chi$ and $\chi^\dagger$ in the matrix element.  Assuming that the
sum over states is dominated by the vacuum, we have
\begin{equation}
\langle \eta_c | \psi^\dagger \chi\chi^\dagger \psi|\eta_c\rangle
\;\approx\; \left| \langle 0 | \chi^\dagger \psi | \eta_c \rangle \right|^2.
\label{VSA-eta}
\end{equation}
In the color-singlet model, the $\eta_c$-to-vacuum matrix element 
on the right side of (\ref{VSA-eta}) can be expressed in terms of 
the wavefunction:
\begin{equation}
\langle 0 | \chi^\dagger \psi|\eta_c \rangle
\;\approx\; \sqrt{2 M_{\eta_c}} \, \sqrt{3 \over 2 \pi} R(0) .
\label{wf-eta}
\end{equation}
Inserting (\ref{VSA-eta}) into (\ref{Gam-eta})
and identifying $M_{\eta_c}$ with $2 m_c$, we
reproduce the result (\ref{Gam-eta:4}) from the color-singlet model.

\subsection{NRQCD Factorization Formula}

In the expression (\ref{Gam-eta}) for the decay rate of the $\eta_c$,
short-distance and long-distance effects have been
factored.  Long-distance effects involving the quarkonium wavefunction appear
only in the NRQCD matrix element, which scales like $M^4v^3$
according to the velocity-scaling rules
in Table~\ref{tab:vscaling}.  Short-distance effects involving the
annihilation of the $Q \overline Q$ pair appear only in the coefficient
$2 {\rm Im} f/M^2$, which is expressed in terms of the fundamental
parameters $\alpha_s$ and $m_c$ of QCD. Thus the expression for the
decay rate in (\ref{Gam-eta}) scales like $\alpha_s^2 v^3$.

The formula (\ref{Gam-eta}) can be generalized to all orders in $\alpha_s$
and to all orders in $v$.  The general factorization formula for the
annihilation decay rate of a quarkonium state $H$ is
\begin{equation}
\Gamma(H)
\;=\; {1 \over 2 M_H}
\sum_{mn} C_{mn} \langle H | {\cal O}_{mn} | H \rangle ,
\label{Gam-fact}
\end{equation}
where the onium state $| H \rangle = | H ({\bf P}=0) \rangle$ 
has the standard relativistic normalization.
The sum in (\ref{Gam-fact})
extends over all operators that can appear in the NRQCD lagrangian
and that have the form
\begin{equation}
{\cal O}_{mn}
\;=\; \psi^\dagger {\cal K}_m \chi \chi^\dagger {\cal K}_n \psi .
\label{O-mn}
\end{equation}
These operators must be gauge
invariant, invariant under parity and charge-conjugation, and scalars under
rotations.  Each of the factors ${\cal K}_n$ and ${\cal K}_m$ 
is the product of a
spin-matrix ($1$ or $\sigma^i$), a color matrix ($1$ or $T^a$), and a
polynomial in ${\bf D}$ and $[D_0, {\bf D}]= g{\bf E}$.  The operator ${\cal
O}_{mn}$ in (\ref{O-mn}) annihilates a $Q \overline Q$ pair in a color and
angular-momentum state determined by ${\cal K}_n$ and creates a $Q \overline
Q$ pair
at the same point in a state determined by ${\cal K}_m$.

The NRQCD factorization formula (\ref {Gam-fact}) untangles the effects of
short distances of order $1/M$ from those of long distances of
order $1/(Mv)$ or larger.  All long-distance effects involving the quarkonium
wavefunction are factored into the NRQCD matrix elements.  Short-distance
effects involving the annihilation of the $Q \overline Q$ pair are contained
in the coefficients.  The coefficient $C_{mn}$ in (\ref{Gam-fact}) is twice
the imaginary part of the coefficient of the operator ${\cal O}_{mn}$ in the
NRQCD lagrangian.  If that operator has scaling dimension $d_{mn}$, then
$C_{mn}$ is $1/m_Q^{d_{mn}-4}$ multiplied by a power series in $\alpha_s(m_Q)$.

The coefficients $C_{mn}$ can be calculated using perturbative matching
methods.  A general matching prescription,
called the {\it threshold expansion method},
has been developed by Braaten and Chen.\cite{Braaten-Chen}
The matching calculations are carried out using
perturbative asymptotic states
$c \bar c = c \bar c({\bf q},\xi,\eta)$ that consist
of a $c$ and a $\bar c$ with relative momentum ${\bf q}$ and in a spin/color
state that is represented by the Pauli spinors $\xi$ and $\eta$.
The matching prescription is
\begin{eqnarray}
&&\sum_X \; (2 \pi)^4 \delta^4(P - k_X) \;
        ({\cal T}_{c \bar c' \to  X})^*
        {\cal T}_{c \bar c \to  X} \Big|_{pQCD}
\nonumber \\
&& \hspace{1in} \;\approx\; \sum_{m n}
C_{mn}\; \langle c \bar c' | \psi^\dagger {\cal K}_m \chi  \;
        \chi^\dagger {\cal K}_n \psi | c \bar c \rangle \Big|_{pNRQCD} \,,
\label{TT-match}
\end{eqnarray}
where $P = (2 \sqrt{m_c^2 + {\bf q}^2}, {\bf 0})$ is the four-momentum
of the $c \bar c$ pair and
${\cal T}_{c \bar c \to  X}$ is the T-matrix element for its
annihilation into a final state $X$ consisting
of light partons.  The sum over $X$ on the left side of
(\ref{TT-match}) includes integration over the phase space of the light
partons.
The complete determination of the short-distance coefficients
requires the use of
different states $c \bar c$ and $c \bar c'$ in the T-matrix element
and in its complex conjugate.
In the matching procedure, the left side of (\ref{TT-match}) is
calculated using perturbative
QCD, and then expanded in powers of the relative momenta
${\bf q}$ and ${\bf q}'$.
The matrix elements on the right side are calculated using
perturbative NRQCD, and then expanded in powers of ${\bf q}$ and ${\bf q}'$.
The coefficients $C_{mn}$ are determined by matching these
expansions order by order in $\alpha_s$.

The relative importance of the various terms in the factorization formula
(\ref{Gam-fact}) is determined by the order in $\alpha_s$ of the coefficient
$C_{mn}$ and by the order in $v$ of the matrix element.  The magnitude of the
matrix element can be estimated using the velocity-scaling rules for operators
given in Table~\ref{tab:vscaling} and the estimates for the probabilities of
higher Fock states in Coulomb gauge given in Section~\ref{sec:Fock}.  
If the operator ${\cal O}_{mn}$
annihilates and creates a $Q\overline Q$ pair in the same color and
angular-momentum state as in the dominant $|Q \overline Q \rangle$ Fock state
of $H$, then the estimate for the matrix element is obtained by 
dividing the estimate for the operator ${\cal O}_{mn}$ 
from Table~\ref{tab:vscaling} by $M^2v^3$.  Otherwise, 
we must take into account suppression factors from the
transitions required to go from the dominant $|Q \overline Q \rangle$ state
to a Fock state in which the $Q \overline Q$ pair can be annihilated by the
operator ${\cal O}_{mn}$ and then back to the
dominant  $|Q \overline Q \rangle$ state.  
There is a suppression factor of $v$ for every electric
transition and a suppression factor of $v^{3/2}$ for every magnetic
transition.  Let $E$ and $M$ be the total number of electric and magnetic
transitions required.  If ${\cal K}_m$ and ${\cal K}_n$ contain $D$ factors of
the covariant derivative ${\bf D}$
but no factors of $g{\bf E}$ or $g {\bf B}$, then the matrix
element $\langle H|{\cal O}_{mn}|H\rangle$ scales like
$v^{3 + D + E + 3M/2}$.  For each factor of $g{\bf E}$ or $g {\bf B}$,
there is an additional suppression factor of $v^3$ or $v^4$, respectively.

Spin symmetry relates NRQCD matrix elements for quarkonium states
that differ only in their spin quantum numbers.  An example involving  
the $\eta_c$ and $J/\psi$ (the lowest $^1S_0$ and $^3S_1$ states of
charmonium) is
\begin{equation}
\langle J/\psi | \psi^\dagger \mbox{\boldmath $\sigma$} \chi \cdot
        \chi^\dagger \mbox{\boldmath $\sigma$} \psi | J/\psi \rangle
\;\approx\;
\langle \eta_c | \psi^\dagger \chi \chi^\dagger \psi | \eta_c \rangle.
\label{HQSS}
\end{equation}
Since spin symmetry is only an approximate symmetry of NRQCD
that is broken at order
$v^2$, the equality (\ref{HQSS}) holds only up to
corrections of relative order $v^2$.

The vacuum-saturation approximation can
be used to express some of the matrix elements in the factorization formula
(\ref{Gam-fact}) in terms of vacuum-to-quarkonium matrix elements.  This
approximation can only be applied if the operator ${\cal O}_{mn}$ annihilates
and creates $Q \overline Q$ pairs in the same color and
angular-momentum state as in the dominant $|Q \overline Q \rangle$ Fock state
of $H$.  In this case, we can insert a complete set of states between $\chi$
and $\chi^\dagger$:
\begin{equation}
\langle H | \psi^\dagger {\cal K}_m \chi
        \chi^\dagger {\cal K}_n \psi | H \rangle
\;=\; \sum_X
\langle H | \psi^\dagger {\cal K}_m \chi | X \rangle
        \langle X | \chi^\dagger {\cal K}_n \psi | H \rangle .
\label{sum-states}
\end{equation}
The vacuum-saturation approximation consists of keeping only the vacuum term
$| 0 \rangle \langle 0 |$ in the sum over states, 
as illustrated in (\ref{VSA-eta}).  The vacuum-saturation
approximation is a controlled approximation 
with an error of relative order $v^4$.  This follows from the fact that in
Coulomb gauge, the next most important term in the sum over states in
(\ref{sum-states}) is a $ | g g \rangle$ Fock state, which
contains two dynamical gluons.  The leading contributions to 
the matrix elements in (\ref{sum-states}) then come from 
$| Q \overline Q g g \rangle$ Fock states whose probabilities are of 
order $v^4$.

The NRQCD matrix elements are sensitive to long-distance 
effects, and therefore can only be calculated using nonperturbative
methods.  The only practical nonperturbative method that is presently
available is Monte Carlo simulations of lattice NRQCD.  
The first such calculations have been carried out recently by
Bodwin, Sinclair, and Kim.\cite{B-K-S}
They demonstrated that the relation (\ref{VSA-eta}) implied
by the vacuum-saturation approximation holds to within numerical accuracy.
They calculated the matrix elements
$\langle 0 | \chi^\dagger \psi | \eta_c \rangle$ and
$\langle 0 | \chi^\dagger {\bf D}^2 \psi | \eta_c \rangle$
and their analogues for bottomonium.
These matrix elements contribute to decays of the lowest S-wave states
at leading order in $v$ and at relative order $v^2$.
They also calculated 
$\langle 0 | \chi^\dagger {\bf D}
        \cdot \mbox{\boldmath $\sigma$} \psi | \chi_{c0} \rangle$
and         
$\langle \chi_{c0} | \psi^\dagger  \mbox{\boldmath $\sigma$} T^a  \chi
        \cdot
\chi^\dagger \mbox{\boldmath $\sigma$} T^a \psi | \chi_{c0} \rangle$
and the analogues of these matrix elements for bottomonium.
As discussed in Section~\ref{sec:Pwaves}, these are the matrix elements 
that contribute to decays  of the lowest P-wave states
at leading order in $v$.
Thus far, the calculations of NRQCD matrix elements have been carried 
out only in minimal NRQCD and without dynamical quarks.

\subsection{Annihilation Decays of P-wave States}
\label{sec:Pwaves}

The NRQCD factorization formula for annihilation decay rates 
has dramatic implications for the decays of $P$-wave states, 
such as $h_c$ (the
$^1P_1$ state of charmonium) and $\chi_{cJ}$, $J=0, 1, 2$ (the $^3P_J$
states).  Calculations of their annihilation decay rates in the color-singlet
model suffer from infrared divergences.  The NRQCD factorization approach not
only resolves the problem of the infrared divergences, but it also leads to
new qualitative insights about $P$-wave charmonium.

We first consider the Fock state expansion of $h_c$ and $\chi_{cJ}$ in Coulomb
gauge.  The dominant $|c \bar c \rangle$ Fock state consists of a
color-singlet $c \bar c$ pair in a $^1P_1$ state for $h_c$ and a $^3P_J$
state for $\chi_{cJ}$.  An electric transition from the dominant Fock state
produces a $|c \bar c g \rangle$ state, with the $c \bar c$ pair in
a color-octet state with angular momentum quantum numbers $^1S_0$ or $^1D_2$
for $h_c$ and $^3S_1$, $^3D_1$, $^3D_2$, or $^3D_3$ for $\chi_{cJ}$.
Therefore these Fock states have probabilities of order $v^2$.  A magnetic
transition from the dominant $|c \bar c \rangle$ Fock state produces a
$|c \bar c g \rangle$ state, with the $c \bar c$ pair in a
color-octet state with angular-momentum quantum numbers $^3P_J$ for $h_c$ and
$^1P_1$ for $\chi_{cJ}$.  This Fock state has a probability of order $v^3$.
All other Fock states have probabilities of order $v^4$ or smaller.

We proceed to identify the most important matrix elements in the NRQCD
factorization formula.  The lowest dimension operator that can
annihilate the $c \bar c$ pair in the  dominant Fock state
of the $h_c$ is
$\psi^\dagger \boldtensorD \chi \cdot \chi^\dagger \boldtensorD \psi$,
where $\boldtensorD$ is defined by
$\chi^\dagger \boldtensorD \psi
        \equiv \chi^\dagger ({\bf D} \psi) - ({\bf D} \chi)^\dagger \psi$.
For $\chi_{cJ}$, the lowest dimension operator has the form
$\psi^\dagger \tensorD {}^m \sigma^n \chi \chi^\dagger
\tensorD {}^i \sigma^j \psi$, where the indices are contracted in different 
ways  for $J=0, 1$, and $2$.  
The matrix elements of these operators for $h_c,
\chi_{c0}, \chi_{c1}$, and $\chi_{c2}$ are related by spin
symmetry.  Up to corrections of
relative order $v^2$, they satisfy
\begin{eqnarray}
&&
\langle h_c | \psi^\dagger  (-\mbox{$\frac{i}{2}$} \boldtensorD) \chi
        \cdot
\chi^\dagger (-\mbox{$\frac{i}{2}$} \boldtensorD) \psi | h_c \rangle
\nonumber \\
&& \hspace{.5in}
\;\approx\; {1 \over 3}
\langle \chi_{c0} | \psi^\dagger (-\mbox{$\frac{i}{2}$} \boldtensorD
                \cdot \, \mbox{\boldmath $\sigma$}) \chi
\chi^\dagger (-\mbox{$\frac{i}{2}$} \boldtensorD
                \cdot \, \mbox{\boldmath $\sigma$}) \psi | \chi_{c0} \rangle
\nonumber \\
&& \hspace{.5in}
\;\approx\; {1 \over 2}
\langle \chi_{c1} | \psi^\dagger (-\mbox{$\frac{i}{2}$} \boldtensorD
                 \times \mbox{\boldmath $\sigma$}) \chi \cdot
\chi^\dagger (-\mbox{$\frac{i}{2}$} \boldtensorD
                \times \mbox{\boldmath $\sigma$}) \psi | \chi_{c1} \rangle
\nonumber \\
&& \hspace{.5in}
\;\approx\;
\langle \chi_{c2} | \psi^\dagger (-\mbox{$\frac{i}{2}$} \tensorD)^{(m}
                 \sigma^{n)} \chi
\chi^\dagger (-\mbox{$\frac{i}{2}$} \tensorD)^{(m}
                 \sigma^{n)}  \psi | \chi_{c2} \rangle \,,
\label{matel-Pwave}
\end{eqnarray}
where $T^{(mn)}$ denotes the symmetric traceless part of a tensor $T^{mn}$.
The vacuum-saturation approximation can be used to express these
matrix element in a simpler form.  Up to corrections of relative order $v^4$,
the matrix element for the  $\chi_{c0}$ can be written
\begin{equation}
\langle \chi_{c0} | \psi^\dagger  (-\mbox{$\frac{i}{2}$} \boldtensorD
        \cdot \, \mbox{\boldmath $\sigma$}) \chi
\chi^\dagger (-\mbox{$\frac{i}{2}$} \boldtensorD
        \cdot \, \mbox{\boldmath $\sigma$}) \psi | \chi_{c0} \rangle
\;\approx\;
\left| \langle 0 | \chi^\dagger (-\mbox{$\frac{i}{2}$} \boldtensorD
        \cdot \, \mbox{\boldmath $\sigma$}) \psi
        | \chi_{c0} \rangle \right|^2 \,.
\label{VSA-chi}
\end{equation}
In the color-singlet model,
the vacuum-to-$\chi_{c0}$ matrix element on the right side of (\ref{VSA-chi})
can be expressed in terms of the radial wavefunction $R(r)$ for the
$P$-wave states:
\begin{equation}
\langle 0 | \chi^\dagger (-\mbox{$\frac{i}{2}$} \boldtensorD
        \cdot \, \mbox{\boldmath $\sigma$}) \psi
        | \chi_{c0} \rangle
\;\approx\; \sqrt{2 M_{\chi_{c0}}} \sqrt{ 9 \over 2 \pi} R'(0) .
\label{wf-h}
\end{equation}

According to the velocity-scaling rules, the matrix elements in
(\ref{matel-Pwave}) scale like $v^5$.  
For a consistent analysis, we must also include all other
matrix elements that scale like $v^5$.  By enumerating the possibilities, one
can see that the only other operators whose matrix elements scale like $v^5$
are $\psi^\dagger T^a \chi \chi^\dagger T^a \psi$ for $h_c$ and
$\psi^\dagger \mbox{\boldmath $\sigma$} T^a \chi \cdot
        \chi^\dagger \mbox{\boldmath $\sigma$} T^a \psi$ for $\chi_{cJ}$.
These operators annihilate and create $c \bar c$ pairs in color-octet
$^1S_0$ and $^3S_1$ states, respectively.  Spin symmetry implies that the
matrix elements for $h_c, \chi_{c0}$, $\chi_{c1}$, and $\chi_{2}$ are equal,
up to corrections of relative order $v^2$:
\begin{equation}
\langle h_c | \psi^\dagger  T^a \chi \chi^\dagger T^a \psi | h_c \rangle
\;\approx\;
\langle \chi_{cJ} | \psi^\dagger  \mbox{\boldmath $\sigma$} T^a  \chi
        \cdot
\chi^\dagger \mbox{\boldmath $\sigma$} T^a \psi | \chi_{cJ} \rangle ,
\qquad J = 0,1,2.
\end{equation}

We have found that,
up to corrections that are suppressed by $v^2$,  
the annihilation decay rates of the P-wave states
can all be expressed in terms of the following two
independent matrix elements:
\begin{eqnarray}
\langle {\cal O}_1 \rangle &\equiv&
{ \left| \langle 0 | \chi^\dagger (-\mbox{$\frac{i}{2}$} \boldtensorD
        \cdot \, \mbox{\boldmath $\sigma$}) \psi | \chi_{c0} \rangle \right|^2
        \over  2 M_{\chi_{c0}} } \,,
\\
\langle {\cal O}_8 \rangle  &\equiv&
{\langle \chi_{c0} | \psi^\dagger  \mbox{\boldmath $\sigma$} T^a  \chi
        \cdot
\chi^\dagger \mbox{\boldmath $\sigma$} T^a \psi | \chi_{c0} \rangle
        \over  2 M_{\chi_{c0}} } \,.
\end{eqnarray}
Their short-distance coefficients can be calculated as power series in
$\alpha_s(m_c)$.  The annihilation processes that contribute to the
coefficients at order $\alpha_s^2$ are $c \bar c \to g g $ and $c
\overline c \to q \overline q$.  If we keep only those terms in the
short-distance coefficients, the annihilation decay rate are
\begin{eqnarray}
\Gamma(h_c) &\approx&
{5 \pi \alpha_s^2(m_c) \over 6 m_c^2} \langle {\cal O}_8 \rangle \,,
\\
\Gamma(\chi_{c0}) &\approx&
{4 \pi \alpha_s^2(m_c) \over m_c^4} \langle {\cal O}_1 \rangle
\;+\; {n_f \pi \alpha_s^2(m_c) \over 3 m_c^2} \langle {\cal O}_8 \rangle \,,
\\
\Gamma(\chi_{c1}) &\approx&
{n_f \pi \alpha_s^2(m_c) \over 3 m_c^2} \langle {\cal O}_8 \rangle \,,
\\
\Gamma(\chi_{c2}) &\approx&
{16 \pi \alpha_s^2(m_c) \over 45 m_c^4} \langle {\cal O}_1 \rangle
\;+\; {n_f \pi \alpha_s^2(m_c) \over 3 m_c^2} \langle {\cal O}_8 \rangle \,,
\end{eqnarray}
where $n_f = 3$ is the number  of flavors of light quarks.  
At this order in $\alpha_s$, the decay rates for $h_c$ and 
$\chi_{c1}$ receive contributions from $\langle {\cal O}_8 \rangle$ only,
because Yang's theorem forbids the annihilation process $c \bar c \to g g $
for a $c \bar c$ pair in a state with total angular momentum 1.
At order $\alpha_s^3$, all of the P-wave states have 
contributions from $\langle {\cal O}_1 \rangle$.
At this order in $\alpha_s$,
the short-distance coefficients of $\langle {\cal O}_1 \rangle$
depend logarithmically on a
factorization scale $\mu$ that can be
interpreted as an infrared cutoff on the energy of soft gluons.  The
matrix element $\langle {\cal O}_8 \rangle$ also depends
logarithmically on $\mu$, which in this case can be identified with
the ultraviolet cutoff of NRQCD.  The $\mu$-dependence cancels
between $\langle {\cal O}_8 \rangle$ and the coefficient
of $\langle {\cal O}_1 \rangle$.\cite{B-B-L:1}
In the color-singlet model, the $\langle {\cal O}_8 \rangle$ terms
are absent and the decay rate depends logarithmically on the 
infrared cutoff $\mu$.  The NRQCD factorization approach 
provides a simple and natural solution to this problem.

The NRQCD factorization formula for a $P$-wave state has a simple
interpretation in terms of the Fock state expansion in Coulomb gauge.  The
color-singlet terms proportional to $\langle {\cal O}_1 \rangle$
are contributions from the dominant $|c \bar c
\rangle$ Fock state, while the color-octet terms proportional to
$\langle {\cal O}_8 \rangle$ are contributions from a $|c
\overline c g \rangle$ Fock state.  The $| c \bar c g \rangle$ state has a
small probability of order $v^2$, and its effects on most observables are
small compared to those of the $|c \bar c \rangle$ state.
However, in the case of the annihilation decay rate, the effects of the 
$| c \bar c \rangle$ state are suppressed by $v^2$ due to the orbital angular
momentum of the $| c \bar c \rangle$ pair.  The contribution of the 
$| c \bar c g \rangle$ Fock state 
has no angular-momentum suppression and 
therefore contributes at the same order in $v$.  
For the $h_c$ and $\chi_{c0}$, the effects of the $|c \bar c g \rangle$ 
Fock state actually dominate, because the short-distance
coefficients of $\langle {\cal O}_1 \rangle$ are suppressed by a factor of
$\alpha_s(m_c)$.

\section{Inclusive Production of Heavy Quarkonium}
\label{sec:IncProd}

\subsection{Topological Factorization}
\label{sec:Top-Fac}

The cross section for producing a quarkonium state $H$ in a
high energy process
necessarily involves  both ``short distances'' of order
$1/M$ or smaller and ``long distances''  of order $1/(M v)$
or larger.
The creation of the $Q \overline Q$ pair involves short distances,
because the  parton processes that produce
the $Q \overline Q$ pair always involve particles
that are off their mass shells
by amounts of order $M$ and can therefore propagate only over
short distances.
The binding of the $Q$ and $\overline Q$ into the state $H$
involves long distances, because gluons whose wavelengths are comparable to
or larger than the size of the bound state, which is of
order $1/(M v)$, play a large role in the binding.

The production of quarkonium in a high energy physics
process typically involves another {\it hard} momentum scale $Q$
in addition to the scale $M$.
If the production cross section is sufficiently inclusive, 
it can be described by an NRQCD
factorization formula that separates short-distance effects
involving the momentum scales $Q$ and $M$ from long-distance effects
that involve lower momentum scales.
It is convenient to separate the
derivation of the factorization formula into 
two steps.  In the first step, which we refer to as
{\it topological factorization},
the standard factorization methods of
perturbative QCD are used to separate the effects of the
hard momentum scale $Q$ from those of the {\it soft}
momentum scale $\Lambda_{QCD}$.
These methods are applicable even if we identify the hard scale
$Q$ with the heavy quark mass $M$.  An additional step is required to
separate the effects of the scale $M$ from those of the scale $Mv$.

The general expression for the inclusive cross section for the production
of a quarkonium state $H$ with four-momentum $P$ is
\begin{eqnarray}
\sum_X d \sigma(12 \to H(P) + X) &=&
{1 \over 4 E_1 E_2 v_{12}}\;  {d^3P \over (2 \pi)^3 2 E_P}
\nonumber \\
&& \hspace{-1in} \times
\sum_X \; (2 \pi)^4 \delta^4(k_1 + k_2 - P - k_X) \;
        |{\cal T}_{1 2 \to H(P) + X}|^2 \,,
\label{dsig}
\end{eqnarray}
where ${\cal T}_{1 2 \to H(P) + X}$ is a T-matrix element
for producing $H$ and the additional particles $X$
and the sum on the right side includes integration over the
phase space of the additional particles.
At the parton level, both ${\cal T}_{12 \to H(P)+X}$ and its complex conjugate
can be expressed as sums of Feynman diagrams.  The product of a single diagram
in ${\cal T}_{12 \to H(P)+X}$ and a single diagram in  ${\cal T}^*_{12 \to
H(P)+X}$ is called a ``cut Feynman diagram.''  Using the factorization methods
of perturbative QCD, one can identify the cut diagrams that dominate in the
limit $Q \to \infty$.  
After taking into account cancellations between real and virtual 
soft gluons, the dominant cut diagrams have the following 
structure:
\begin{itemize}

\item a {\it hard-scattering subdiagram} ${\cal H}$ to the left of the cut.  
The outgoing lines include a $Q \overline Q$ pair with small relative 
momentum of order $Mv$ and additional hard partons.  
There can also be incoming hard-parton lines if the process involves 
hadrons in the initial state.

\item a {\it hard-scattering subdiagram} ${\cal H}^*$ to the right of the cut 
that is just the mirror image of ${\cal H}$.

\item a {\it jet-like subdiagram} ${\cal J}_i$ 
for each of the hard partons  
attached to ${\cal H}$.  The subdiagram extends through the cut
and is attached to ${\cal H}$ and to ${\cal H}^*$ by 
single hard parton lines.

\item an {\it onium subdiagram} ${\cal O}$ that extends through the cut
and is attached to ${\cal H}$ and to ${\cal H}^*$  by $Q$ and $\overline Q$ 
lines that have small relative momentum.

\end{itemize}
The cut diagrams that do not have the above structure are suppressed by powers
of $1/Q$.
With this topological factorization of the dominant cut diagrams,
all effects involving the hard momentum scale $Q$ are factored into the 
hard-scattering subdiagrams ${\cal H}$ and ${\cal H}^*$,
while all effects of the soft scale $\Lambda_{QCD}$ are factored into
${\cal O}$, ${\cal J}_1$, ${\cal J}_2$, $\ldots$.
The gluon interactions that bind the $Q \overline Q$ pair
into the onium state $H$ are also contained within the 
onium subdiagram ${\cal O}$.

The proofs of the factorization theorems of perturbative QCD
are very difficult,  and explicit proofs are available only for a 
very few processes, such as inclusive hadron production in $e^+e^-$ 
annihilation and the Drell-Yan process for lepton pair production 
in hadron collisions.\cite{Collins-Soper}  However there is no apparent
obstacle to extending these proofs to inclusive onium production.

\subsection{NRQCD Factorization}
\label{sec:NRQCD-Fac}

After topological factorization, the effects of the scale $Mv$ 
are distributed in a complicated way between ${\cal H}$, 
${\cal H}^*$, ${\cal O}$, and the
$Q$ and $\overline Q$ propagators that connect them.  
The onium subdiagram ${\cal O}$ involves the scale $Mv$ 
because gluons with momentum of order $Mv$
play an important role in the binding of the $Q \overline Q$ 
pair into an onium state.  The hard-scattering subdiagrams 
${\cal H}$ and ${\cal H}^*$ involve the scale $Mv$, 
because the outgoing $Q$ and $\overline Q$ lines have
relative momenta on the order of $Mv$.  It is in factoring 
the scale $Mv$ out of ${\cal H}$ and ${\cal H}^*$ that 
NRQCD enters into the picture.

Consider the part of the cut diagram that includes ${\cal H}$, 
${\cal O}$, and the $Q$ and $\overline 	Q$ propagators that connect them.  
If the $Q$ and $\overline Q$ have four-momenta ${1 \over 2} P + q$ 
and ${1 \over 2} P - q$, then the diagram
also involves an integral over the relative momentum $q$.  A simple way to
disentangle the momentum scale $Mv$ from ${\cal H}$ is to expand it as a
Taylor series in $q$ and absorb the factors of $q$ as well as the integration
over $q$ into ${\cal O}$.  
Each term in the Taylor expansion corresponds to a local operator that
creates a $Q \overline Q$ pair from the vacuum.
By applying a similar procedure to disentangle the momentum scale $Mv$ 
from ${\cal H}^*$, the onium subdiagram is reduced to 
vacuum-expectation values of local operators that create and annihilate  
$Q \overline Q$ pairs.  After a renormalization group transformation
and appropriate field redefinitions, the matrix elements can be 
expressed as expectation values in the NRQCD vacuum of the form
\begin{equation}
\langle {\cal O}^H_{mn} \rangle \;=\; 
\langle 0 | \chi^\dagger {\cal K}_m \psi \; {\cal P}_H \;
	 \psi^\dagger {\cal K}_n \chi | 0 \rangle \,, 
\label{O-H}
\end{equation}
where ${\cal P}_H$ projects onto states that in the asymptotic future
contain the quarkonium state $H$ plus soft partons $S$
whose total energy is
less than  the ultraviolet cutoff of NRQCD:
\begin{equation}
{\cal P}_H \;=\;
\sum_S | H + S \rangle \langle H + S |  \,. 
\label{P-H}
\end{equation}

If we integrate the dominant cut diagrams
over the phase space of all the hard partons in the 
process, we obtain the NRQCD factorization formula for the inclusive
cross section:
\begin{equation}
\sum_X d \sigma(12 \to H(P) + X) \;=\;
{1 \over 4 E_1 E_2 v_{12}}\;  {d^3P \over (2 \pi)^3 2 E_P}
\sum_{mn}  C_{mn}(k_1,k_2, P) \; \langle {\cal O}^H_{mn} \rangle \,. 
\label{dsig-fact}
\end{equation}
The sum in (\ref{dsig-fact}) extends over all NRQCD matrix elements 
of the form (\ref{O-H}).  The product of the operators 
$\chi^\dagger {\cal K}_m \psi$ and $\psi^\dagger {\cal K}_n \chi$
must be gauge-invariant.  It need not be rotationally
invariant if the quarkonium state $H$ is polarized.
In the factorization formula (\ref{dsig-fact}), the 
hard-scattering subdiagrams ${\cal H}$ and ${\cal H}^*$ and the 
jet-like subdiagrams ${\cal J}_i$ have all been subsumed in the 
short-distance coefficients $C_{mn}$. 

Since the coefficients $C_{mn}$ in (\ref{dsig-fact}) 
involve only short distances of order $1/M$ or larger, they can
be expressed as perturbation series in $\alpha_s(M)$.
The {\it threshold expansion method} provides a general prescription 
for calculating the short-distance coefficients.\cite{Braaten-Chen}
Denoting by $c \bar c(P)$  a state consisting of a $c$ and $\bar c$ 
with relative momentum ${\bf q}$
that has been boosted to four-momentum $P$,
the matching prescription is 
\begin{eqnarray}
&&\sum_X \; (2 \pi)^4 \delta^4(k_1 + k_2 - P - k_X) \;
	({\cal T}_{1 2 \to c \bar c'(P) + X})^* 
	{\cal T}_{1 2 \to  c \bar c(P) + X} \Big|_{pQCD} 
\nonumber \\
&& \hspace{.2in}
\;\approx\; \sum_{m n}
C_{mn}(k_1,k_2,P)
	\langle 0 | \chi^\dagger {\cal K}_m \psi  \; 
	{\cal P}_{c \bar c',c \bar c} \;
	\psi^\dagger {\cal K}_n \chi | 0 \rangle \Big|_{pNRQCD} \,, 
\label{TT-match:prod}
\end{eqnarray}
where the projection operator in the NRQCD matrix element is 
\begin{equation}
{\cal P}_{c \bar c',c \bar c} \;=\;
\sum_S |  c \bar c' + S \rangle 
		\langle c \bar c + S |  \,. 
\label{P-ccbar}
\end{equation}
The left side of (\ref{TT-match:prod}) is to be calculated using perturbative
QCD, and then expanded in powers of the relative momenta
${\bf q}$ and ${\bf q}'$ of the $c \bar c$ pairs.
The matrix elements on the right side are to be calculated using 
perturbative NRQCD, and then expanded in powers of ${\bf q}$ and ${\bf q}'$.
The coefficients $C_{mn}$ are then determined by matching these
expansions order by order in $\alpha_s$.

The relative importance of the various terms in the factorization formula
(\ref{Gam-fact}) is determined by the magnitudes of the coefficients $C_{mn}$
and by the order in $v$ of the matrix elements 
$\langle {\cal O}^H_{mn} \rangle$.
The size of the coefficient $C_{mn}$ is determined not only by the order in 
$\alpha_s$, but also by its dependence on  
dimensionless ratios of kinematic variables that are involved in the 
$c \bar c$ production process.  
The magnitudes of the matrix elements can be estimated 
by using the velocity-scaling rules for operators
given in Table~\ref{tab:vscaling} and the scaling with $v$ of 
the rates for electric and magnetic transitions.  
If the operator ${\cal O}^H_{mn}$
creates and annihilates a $Q\overline Q$ pair in the same color and
angular-momentum state as in the dominant $|Q \overline Q \rangle$ Fock state
of $H$, then the magnitude of  the matrix element 
$\langle {\cal O}^H_{mn} \rangle$ is estimated by multiplying the 
factors in Table~\ref{tab:vscaling} and dividing
by $M^2v^3$.  For other matrix elements, we must take into account 
suppression factors from the transitions required to go from
the $Q \overline Q$ state created by the
operator $\psi^\dagger {\cal K}_n \chi$  to a state
in which the $Q \overline Q$ pair has the same quantum numbers as 
in the dominant $|Q \overline Q \rangle$ Fock state and then to a state 
in which the  $Q \overline Q$ pair can be annihilated by the operator
$\chi^\dagger {\cal K}_m \psi$.  There is a suppression factor of $v$ 
for every electric transition that is required 
and a suppression factor of $v^{3/2}$ for every magnetic
transition.  The scaling of the production matrix element
$\langle {\cal O}^H_{mn} \rangle$ with $v$
is identical to that of the corresponding decay matrix
element $\langle H|{\cal O}_{mn}|H\rangle$.

The NRQCD matrix elements that appear in the  NRQCD factorization formula
(\ref{dsig-fact}) can be simplified by using 
symmetries of NRQCD.  
Rotational symmetry is an exact symmetry of NRQCD.  
It implies, for example,  that 
\begin{equation}
\langle \chi^\dagger \sigma^j T^a \psi \; 
		{\cal P}_{J/\psi} 
		\psi^\dagger \sigma^i T^a \chi \rangle \;=\;
{\delta_{ij} \over 3} \langle \chi^\dagger \sigma^k T^a \psi \; 
		{\cal P}_{J/\psi} \;
		\psi^\dagger \sigma^k T^a \chi \rangle .
\label{rot-sym}
\end{equation}
Spin symmetry is an approximate symmetry of NRQCD
that holds up to corrections of order $v^2$.  
It implies, for example,  that 
\begin{equation}
\langle \chi^\dagger \sigma^j T^a \psi \; 
		{\cal P}_{\psi(\lambda)} 
		\psi^\dagger \sigma^i T^a \chi \rangle \;\approx\;
U_{\lambda j} U^\dagger_{i \lambda}
\langle \chi^\dagger T^a \psi \; 
		{\cal P}_{\eta_c} \;
		\psi^\dagger T^a \chi \rangle ,
\label{hqs-sym}
\end{equation}
where $\lambda$ specifies the polarization of the $J/\psi$ 
and $U_{i \lambda}$ is the unitary matrix that transforms vectors 
from the spherical basis to the Cartesian basis. 

The vacuum-saturation approximation can be used to simplify the
matrix elements of operators that create and annihilate 
$c \bar c$ pairs in the dominant Fock state of the quarkonium.
In the vacuum-saturation approximation, the projection operator
$P_H$ defined in (\ref{P-H}) is replaced by 
the single term $ | H \rangle \langle H|$. 
This is a controlled approximation in NRQCD,
holding up to corrections that are of order $v^4$.
It implies, for example, that
\begin{equation}
\langle 0 | \chi^\dagger \sigma^i \psi \; {\cal P}_{\psi(\lambda)} \;
	 \psi^\dagger \sigma^j \chi | 0 \rangle
\;\approx\;
\langle 0 | \chi^\dagger \sigma^i \psi | \psi(\lambda) \rangle  \; 
	\langle \psi(\lambda) | \psi^\dagger  \sigma^j \chi | 0 \rangle \,.
\label{vsa-prod}
\end{equation}

The matrix elements in the NRQCD factorization formula
involve long-distance effects and therefore
can only be calculated using nonperturbative
methods.  Unfortunately, in contrast to the decay matrix elements 
$\langle H | {\cal O}_{mn} | H \rangle$,
there are no effective prescriptions for calculating the 
production matrix elements 
$\langle {\cal O}^H_{mn} \rangle$ using lattice NRQCD.
The problem lies in implementing on the lattice
the projection defined by (\ref{P-H}). 
Thus these NRQCD matrix elements must be treated as phenomenological 
parameters to be determined by experiment.  
The only exceptions are the matrix elements to which the vacuum-saturation 
approximation can be applied.  Vacuum-to-quarkonium matrix elements
of the form $\langle H | \psi^\dagger {\cal K}_n \chi | 0 \rangle$
can be calculated using Monte-Carlo simulations of NRQCD.

\subsection{Prompt Charmonium at the Tevatron}
\label{sec:Prompt}

The NRQCD factorization framework (\ref{dsig-fact})
has many applications, some of which are described in a recent 
review.\cite{B-F-Y}  One application for which the implications  
are particularly dramatic is the production of prompt charmonium
at large transverse momentum in $p \bar p$ collisions. 
At sufficiently large transverse momentum $p_T$,
the cross section for $p \bar p \to \psi + X$
is dominated by gluon fragmentation.\cite{Braaten-Yuan}
It can be factored into
the cross section for producing a gluon with transverse momentum
$p_T/z$ and a fragmentation function $D_{g \to \psi}(z)$ that
gives the probability
that the jet initiated by the gluon includes a $\psi$ carrying
a fraction $z$ of the gluon momentum:
\begin{equation}
d \sigma (p \bar p \to \psi(P) + X) \;=\;
\int_0^1 dz \; d \hat{\sigma} (p \bar p \to g(P/z) + X) \;
        D_{g \to \psi}(z) \,.
\label{dsig-frag}
\end{equation}
Using the NRQCD factorization
approach, the fragmentation function can be expressed in the form
\begin{equation}
D_{g \to \psi}(z) \;=\;
\sum_{mn} d_{mn}(z) \langle {\cal O}_{mn}^\psi \rangle \,,
\label{frag-fact}
\end{equation}
where all effects of the momentum scale $m_c$ have been factored into the 
short-distance coefficients $d_{mn}(z)$.  The relative importance of 
the various terms in the fragmentation function is determined by the order
in $v$ of the matrix element and the order in $\alpha_s$ of its coefficient.

The matrix element that is leading order in $v$ is 
$ \langle \chi^\dagger \sigma^k \psi {\cal P}_\psi 
	\psi^\dagger \sigma^k \chi \rangle$,
which scales like $v^3$.
This term corresponds to the formation of a $\psi$ from a 
$c \bar c$ pair that is created in a color-singlet $^3S_1$ state,
and it is the only contribution in  the color-singlet model. 
The leading contribution to the short-distance coefficient of this term 
in the gluon fragmentation function (\ref{frag-fact}) is of order $\alpha_s^3$
and comes from the parton process $g^* \to c \bar c g g$.
Keeping only this term in the fragmentation function (\ref{frag-fact}),
the cross section predicted by (\ref{dsig-frag}) is about a factor of
30 below recent data on prompt $\psi$ production at the Tevatron.

The color-singlet-model term in the gluon fragmentation function 
scales like $\alpha_s^3 v^3$.  All other terms have
matrix elements that scale like $v^5$ or smaller.
There are however terms whose short-distance coefficients  
are suppressed by fewer 
powers of $\alpha_s$.  There is one coefficient in particular 
that is of order $\alpha_s$, because it receives a
contribution from the parton process $g^* \to c \bar c$. 
The matrix element is 
$\langle \chi^\dagger \sigma^k T^a \psi  {\cal P}_\psi 
         \psi^\dagger   \sigma^k T^a \chi \rangle$
and it scales like $v^7$.
This term corresponds to the formation of a $\psi$ from a 
$c \bar c$ pair that is created in a color-octet $^3S_1$ state.  
At leading order in $\alpha_s$, this term in the fragmentation function is
\begin{equation}
D_{g \to \psi}(z) \;=\;
{\pi \alpha_s(m_c) \over 96 m_c^4} \delta(1-z) \;
\langle \chi^\dagger \sigma^k T^a \psi \; {\cal P}_\psi \;
         \psi^\dagger   \sigma^k T^a \chi \rangle \,.
\label{D-psi}
\end{equation}
Braaten and Fleming proposed that the enhancement from the two 
fewer powers of $\alpha_s$ relative to the color-singlet model term
can overcome the suppression by $v^4$, and that this term 
might dominate the gluon fragmentation function.\cite{Braaten-Fleming}  
The $p_T$-dependence predicted by this  mechanism is in agreement
with the Tevatron data.  The normalization depends on the unknown matrix element
in (\ref{D-psi}), but the value of
the matrix element required to fit the data is consistent with
suppression by a factor of $v^4$ relative to the 
color-singlet-model matrix element
$\langle \chi^\dagger \sigma^k \psi {\cal P}_\psi 
	\psi^\dagger \sigma^k \chi \rangle$.

Cho and Wise pointed out that this production mechanism has dramatic 
implications for the polarization of the $\psi$.\cite{Cho-Wise}
At leading order in $\alpha_s$, the $\psi$'s produced by gluon
fragmentation will be 100\% transversely polarized.
The radiative corrections to the fragmentration function 
were examined by Beneke and Rothstein, 
and they concluded that the spin alignment
at large $p_T$ will remain greater than 90\%.\cite{Beneke-Rothstein}
The largest corrections to the spin alignment at values of $p_T$
that can be measured at the Tevatron come from
nonfragmentation contributions that fall like $1/p_T^2$, 
and these contributions have recently been calculated by  
Beneke and Kraemer.\cite{Beneke-Kramer}
An experimental measurement of the spin alignment
in agreement with these predictions would constitute a dramatic triumph
of the NRQCD factorization approach.

\section{Conclusions}
\label{sec:Conc}

The NRQCD factorization formulas (\ref{Gam-fact}) and (\ref{dsig-fact})
provide a firm theoretical foundation for analyzing
annihilation decay rates and inclusive production rates of heavy quarkonium.
The short-distance coefficients can be calculated as power series in the
running coupling constant $\alpha_s(M)$,
and the long-distance factors are defined in terms of NRQCD matrix elements
that scale in a definite way with $v$.
This approach not only provides a framework for
carrying out systematic quantitative calculations of
quarkonium processes, but it also leads to new qualitative
insights into quarkonium physics.

\section*{Acknowledgments}

This work was supported in part by the U.S.
Department of Energy, Division of High Energy Physics, under
Grant DE-FG02-91-ER40684.  I thank G.T. Bodwin and G.P. Lepage
for valuable discussions and I. Maksymyk for a careful and critical
reading of the manuscript.  I would also like to thank the organizers of the
Third International Workshop on Particle Physics Phenomenology
for their hospitality.

\end{document}